\documentclass[pra,twocolumn,amsmath,amssymb,superscriptaddress]{revtex4-1}
\usepackage{epsfig,graphicx,graphics,amsmath,amssymb,float}
\usepackage[T1]{fontenc}
\usepackage[latin9]{inputenc}
\usepackage{amsmath}
\usepackage{amssymb}
\usepackage{xcolor}
\usepackage{amscd}
\usepackage{bm}
\usepackage{bbold}
\usepackage{psfrag}
\usepackage{mathrsfs}
\usepackage{float}
\usepackage{textgreek}

\usepackage{bbm} 
\usepackage{chemformula}
\usepackage[caption=false]{subfig}
\usepackage{subfig}
\usepackage{color}
\usepackage[bookmarks=true,colorlinks,linkcolor=blue,urlcolor=blue,citecolor=blue]{hyperref}

\usepackage{dutchcal}
\DeclareMathAlphabet{\matholdcal}{OMS}{cmsy}{m}{n}

\DeclareMathOperator{\sgn}{sgn} 

\newcommand{\be}{\begin{equation}}
\newcommand{\ee}{\end{equation}}
\newcommand{\bea}{\begin{eqnarray}}
\newcommand{\eea}{\end{eqnarray}}

\newcommand{\mc}{\matholdcal}

\newcommand{\te}{\text}

\begin{document}

\title{Edge State Effects in Quantum Work Statistics}

\author{Moallison F. Cavalcante}
\email{moall@unicamp.br}
\affiliation{
Gleb Wataghin Physics Institute, The University of Campinas, 13083-859, Campinas, S\~{a}o Paulo, Brazil}
\affiliation{Department of Physics, University of Maryland, Baltimore County, Baltimore, MD 21250, USA}
\affiliation{Quantum Science Institute, University of Maryland, Baltimore County, Baltimore, MD 21250, USA}
\date{\today}

\begin{abstract}

Motivated by the objective of quantifying the energetic cost of accessing boundary phases through local control, we investigate here a simple, analytically tractable quantum impurity model. This model exhibits a rich boundary phase diagram, characterized by phases with different numbers of edge states. By considering a local quench protocol that drives the system out of equilibrium, we calculate exactly the resulting quantum work distribution across these phases. Our results show that the presence of edge states strongly alters this distribution. 
In particular, we analytically determine the signatures of these states in the low, intermediate, and high-energy sectors.


\end{abstract}


\maketitle
\section{Introduction}
\label{introduction}

Understanding the nonequilibrium behavior of closed quantum many-body systems is a central and challenging problem for both the statistical physics and condensed matter communities \cite{RMP_2011,Eisert2015,Gogolin_2016,Mori_2018,RMP_2019}. The simultaneous presence of strong interactions among many degrees of freedom, which prevents the use of perturbative methods, together with the external driving that steers the system away from an equilibrium state poses significant theoretical difficulties. Therefore, analytical results in such nontrivial regimes are rare and highly valuable, as they shed light on the underlying physical mechanisms and guide the understanding of more complex scenarios.

Currently, several experimental platforms enable highly controlled investigations of a wide variety of physical systems, including bosonic \cite{RMP_Bloch,RMP_Cazalilla_2011,Trotzky2012,Aimet2025}, fermionic \cite{doi:10.1126/science.aaf5134,Evered2025}, and spins systems \cite{Rydn_exp_2021,Dumitrescu2022}, under different nonequilibrium protocols, such as quantum quenches \cite{Mitra_Quench_2018}. These developments have made it possible to address fundamental questions in the field, for instance, equilibration and thermalization \cite{Kinoshita2006,Trotzky2012,Pre_Science_2012,Pre_Science_2016}, as well as the spreading of correlations \cite{Cheneau_2012,Langen2013,w1cp-l5vq}. From a quantum thermodynamics perspective, a recent experiment with ultracold gases \cite{Aimet2025} has also probed Landauer's principle, namely, the minimal energetic cost of erasing one bit of information, in a quantum many-body setting.

Assessing the energetic cost associated with manipulating quantum systems is essential for performing quantum technology tasks, such as quantum annealing \cite{Yarkoni_2022}, efficiently, as well as for characterizing general aspects of out of equilibrium systems \cite{RMP_2011}. Along these lines, the fruitful interplay between quantum thermodynamics and condensed matter physics has stimulated intense interest in the statistics of the work done on closed quantum many-body systems over recent years \cite{PRL_Ale_2008,Sindona_2014,PRE_Landi_2016,Goold2018,PRR_Krissia_2020,PRR_Kiely_2023,Gen_LRT_PRL_2024,PRB_Dora_2024,vn83-mt2v,96cm-ktcb}. Despite substantial progress, the role played by intrinsically many-body features, such as strong interactions, disorder, and topology, in shaping quantum work statistics remains not fully understood.


In this work, we are interested into investigate how edge states can influence the work statistics of a one-dimensional quantum many-body system. These states, which are exponentially localized around the edges of the system, usually appear as a consequence of non-homogeneities in real space \cite{PRL_Tony_2007,PRB_main_ref_2016, MULLER2016482,t78z-pyfr} or, more interestingly, from topological properties \cite{Kitaev_2001,Alicea_2012,CMoore_PRB_2018}. Previous works dedicated to investigate the nonequilibrium physics of these states have demonstrated that they imply slower relaxation towards equilibrium \cite{GarciaRodriguez2018,PRB_count_edge_2023} and trapping of information \cite{OTOC_inf_trap_PRB_2023,Muruganandam_2024,Sur_2024,Moa_PRR_2025}. 

Concretely, we consider the one-dimensional transverse-field Ising model with open boundary conditions in the presence of an edge impurity. The equilibrium ground state of this system hosts up to two edge states in addition to a continuum of bulk states \cite{PRB_main_ref_2016}. The system is prepared in this ground state and then driven out of equilibrium by a sudden local magnetic field, in the form of a ``$\delta(t)$-kick'', acting at the impurity. Using a two-time energy measurement (TTM) scheme, we compute the full work statistics \textit{exactly} and then identify key signatures, such as edge singularities with different exponents, coming from the edge states and the associated nontrivial boundary physics.

This work is organized as follows. In Section \ref{sec_model}, we introduce the model and explain its boundary phase diagram. In Section \ref{sec_quench_pro}, we discuss the quench protocol and the consequent nonequilibrium system state that follows from it. In Section \ref{sec_work_st}, we calculate the work distribution. In Section \ref{sec_res}, analytical and numerical results across the different boundary phases are shown and discussed. In Section \ref{sec_conclu}, our concluding remarks
can be found. Finally, in Appendix \ref{Appen1}, we extend the analysis for quenches of finite duration.

\section{Quantum Impurity model}
\label{sec_model}

This paper aims to study the nonequilibrium physics of the following open boundary condition (OBC) spin-$1/2$ Ising-like chain
\begin{eqnarray}   
 H_{0}&=&-Jh\mu \sigma^z_1- \:Jh\sum_{j=2}^{N}\sigma^z_{j}-J\sum_{j=1}^{N-1}\sigma^x_{j}\sigma^x_{j+1}.
\label{main_Hamil}    
\end{eqnarray}
Here, $\sigma^{a=x,y,z}_j$ are the Pauli matrices at the site $j$, $J\equiv 1$ is the exchange coupling (which will fix the energy scale of the problem) and $h>0$ is the $z$-direction external field magnitude. The first site of the chain describes the impurity (or defect) which here we model assuming a $\mu\neq 1$. The model in Eq. (\ref{main_Hamil}) still undergoes a quantum phase transition between a ferromagnetic ($h<h_c$) and a paramagnetic ($h>h_c$) phase, where the quantum critical point is located at $h_c\equiv 1$ \cite{PRB_main_ref_2016}. In what follows, we will work in the limit of a semi-infinite chain, $N\to \infty$. 

The model in Eq. (\ref{main_Hamil}) was analytically solved in reference \cite{PRB_main_ref_2016}. The solution, which assumes $N\to\infty$, shows that the presence of the edge impurity, $\mu\neq 1$, produces a new fermionic mode $\gamma_2$ localized around the edge $j\sim 1$, beyond the well-studied nonlocal and topologically protected fermionic mode $\gamma_1$ found by Kitaev \cite{Kitaev_2001} in the impurity-free case $\mu=1$. Before discussing the boundary phase diagram established in \cite{PRB_main_ref_2016}, we can qualitatively understand the emergence of these two localized edge modes as follows. 

Decomposing the spin operators $\sigma^a_j$ in terms of real (Majorana) fermions $\eta_{Aj}$ and $\eta_{Bj}$ through a Jordan-Wigner transformation \cite{sachdev1999quantum},
\begin{eqnarray}
    \sigma^x_j&=&-\left( \prod_{n=1}^{j-1}i\eta_{An}\eta_{Bn} \right)\eta_{Aj},\\
    \sigma^z_j&=&i\eta_{Aj}\eta_{Bj},
    \label{JW}
\end{eqnarray}
where $\eta^2_{Aj}=\eta^2_{Bj}=1$, $\{\eta_{Aj},\eta_{Aj'}\}=\{\eta_{Bj},\eta_{Bj'}\}=2\delta_{jj'}$ and $\{\eta_{Aj},\eta_{Bj'}\}=0$, the Hamiltonian of Eq.~(\ref{main_Hamil}) is recast as (apart from a constant)
\begin{equation}
    H_0=iJh\mu \eta_{B1}\eta_{A1}+iJh\sum_{j=2}^{N}\eta_{Bj}\eta_{Aj}-iJ\sum_{j=1}^{N-1}\eta_{Bj}\eta_{Aj+1}.
    \label{H0_Majorana}
\end{equation}
We note that $H_0$ is quadratic and can therefore be directly diagonalized (numerically) for finite $N$. 

First, let us consider the absence of impurity $\mu=1$. For $h=0$, we see that the fermions $\eta_{A1}$ and $\eta_{BN}$ totally decouple from the system. These two real fermions live at the two edges of the chain and combine to form the complex fermionic mode mentioned above, $\gamma_1\equiv\frac{\eta_{A1}-i\eta_{BN}}{2}$. As found by Kitaev \cite{Kitaev_2001}, this nonlocal mode is topologically protected and survives not only in this particular $h=0$ limit but also in the whole ferromagnetic phase $h<1$.

The second localized edge mode $\gamma_2$ appears when we consider the presence of the impurity $\mu\neq1$. For $\mu=0$ and $h\gg 1$, the real fermion $\eta_{A1}$ is fully decoupled, whereas $\eta_{B1}$ remains only weakly hybridized with the chain. These two fermions together constitute the mode $\gamma_2$. A similar situation occurs for $\mu\gg 1$, where now $\eta_{A1}$ and $\eta_{B1}$ are strongly bound to each other but only weakly coupled to the rest of the system. In both regimes, the $\gamma_2$ mode remains entirely localized around the edge $j=1$, highlighting its nontopological origin \cite{CMoore_PRB_2018}.

The qualitative picture described above is supported by the exact solution of the model \cite{PRB_main_ref_2016}. The mode $\gamma_1$ is present for \textit{any} $\mu$ in the ferromagnetic side $h<1$ thanks to its topological protection, which ensures its stability against local perturbations (as the presence of impurities). On the other hand, the mode $\gamma_2$ appears in two distinct regions: it exists for $|\mu| < \sqrt{1 - 1/h}$ when $h > 1$, and for $|\mu| > \sqrt{1 + 1/h}$ for any value of $h$. These different regions where the modes $\gamma_{1,2}$ can exist determine different boundary phases of the model (\ref{main_Hamil}). Finally, for all $(\mu, h)$, a continuum of delocalized (bulk) modes is present. These modes become gapless on the quantum critical line $h_c = 1$ (independent of $\mu$), while they are fully gapped away from this line.

In its diagonal form, $H_0$ becomes
\begin{eqnarray}
    H_0&=&\sum_{\kappa}\Gamma_{\kappa}\gamma^{\dagger}_{\kappa}\gamma_{\kappa},
    \label{H_diag}
\end{eqnarray}
where $\kappa$ accounts for both the bulk ($\kappa=k$) and edge ($\kappa=1,2$) modes with energy $\Gamma_{\kappa}$, and we drop the energy of the ground state in the above equation. The complex fermions $\gamma_{\kappa}$ and the Majorana fermions in Eq. (\ref{H0_Majorana}) are related through
\begin{eqnarray}
    \eta_{Aj}&=&\sum_{\kappa}\phi_{j\kappa}\alpha_{\kappa},\\
    \eta_{Bj}&=&\sum_{\kappa}\psi_{j\kappa}\beta_{\kappa},
\end{eqnarray}
where $\alpha_{\kappa}=\gamma^{\dagger}_\kappa+\gamma_\kappa$ and $\beta_\kappa=i(\gamma^{\dagger}_\kappa-\gamma_\kappa)$, and $\phi_{j\kappa}$ and $\psi_{j\kappa}$ are the respective wave-functions. We refer the reader to Ref. \cite{PRB_main_ref_2016} to find explicit expressions for $\Gamma_\kappa$, $\phi_{j\kappa}$ and $\psi_{j\kappa}$.

In the next section, we will explore nonequilibrium signatures of the different boundary phases discussed above. 

\section{Local quench protocol}
\label{sec_quench_pro}

The protocol chosen here to bring the system out of equilibrium is one that allows us to directly probe the different boundary phases of the model (\ref{main_Hamil}). Initially, for times $t<0$, the system is supposed to be in its ground-state $|\Psi_0\rangle=|0\rangle$. At time $t=0$, a magnetic field is abruptly turned on and off, i.e., in the quench form, at the impurity site. This $\delta$-kick is responsible for injecting energy into the system through the edge, or, in other words, for performing work on the system. 

The system will be governed by the follow time-dependent Hamiltonian  
\begin{equation}
    \mc H(t) = H_0 + (-g_0)\delta(t)\sigma^z_1,
    \label{protocol}
\end{equation}
where $g_0$ describes the intensity of the perturbation and $\delta(t)$ is the Dirac delta function. The system follows a unitary dynamics, so its state is given by $|\Psi(t)\rangle=e^{-iH_0t}\mc U(t)|\Psi_0\rangle$, where $\mc U(t)=\mc T \te{exp}\left(ig_0\int_0^tdt'\:\delta(t')\sigma^z_1(t') \right)=e^{ig_0\sigma^z_1}$. Thus, the many-body nonequilibrium post-quench state reads exactly 
\begin{eqnarray}
    |\Psi(t>0)\rangle&=&\left(\cos\left(g_0\right) + i\sin \left(g_0\right) e^{-iH_0t}\sigma^z_1  \right)|\Psi_0\rangle,\nonumber\\
    \label{state}
\end{eqnarray}
where we set $H_0|0\rangle = E_0|0\rangle\equiv0$, without loss of generality. Given that $\sigma^z_1=i\eta_{A1}\eta_{B1}$ [see Eq. (\ref{JW})], the above state describes a superposition between the vacuum (ground state with zero excitations) and excited states with two excitations. These two excitations will be created in the different modes of $H_0$. 

In what follows, we focus exclusively on the protocol (\ref{protocol}). However, in App. \ref{Appen1}, we extend the analysis for quenches of finite duration $\tau$.

It is worth mentioning that the protocol (\ref{protocol}) was recently used to prepare a \textit{many-body} $X$ state in the phase with two edge states \cite{Moa_PRR_2025}. Such a $X$ state exhibits genuine quantum correlations, such as discord. Therefore, the present work goes in the direction of quantifying the energetic cost of this preparation.

\section{Quantum work statistics}
\label{sec_work_st}

The signatures of the protocol (\ref{protocol}) can be identified through the statistics of quantum work done on the system. We can determine it through a standard two-time projective energy measurement (TTM) scheme \cite{PRE_work_2007,PRE_Talkner_2008,PRL_Ale_2008,Deffner_Campbell_book}. 

Consider the following energy measurement protocol. Before the magnetic field is switched on, we perform the first energy measurement, which yields the ground-state energy $E_0$. The second measurement is carried out after the quench, returning an energy $E_m$ with probability $|\langle \Psi_m | \Psi(t) \rangle|^2$, where $H_0 |\Psi_m\rangle = E_m |\Psi_m\rangle$. The difference $(E_m - E_0)$ is a random variable representing the energy absorbed due to the quench. By repeating this procedure many times, one can construct the quantum work distribution
\begin{equation}
P(w)=\sum_m|\langle\Psi_m|\Psi(t)\rangle|^2\delta\left(w-(E_m-E_0)\right),
\end{equation}
where the sum is over the spectrum of $H_0$. The above equation provides a formal expression for $P(w)$, which in general is highly costly to compute in a many-body context, since it requires knowledge of the exponentially large set $\{E_m,|\Psi_m\rangle\}$. 

Alternatively, the statistics of the work done $P(w)$ can be obtained from the characteristic function $G(t)\equiv\langle e^{iwt} \rangle=\int dw\: e^{iwt}P(w)$ by means of an inverse Fourier transform. It is direct to show that for the protocol in Eq. (\ref{protocol}) we have 
\begin{eqnarray}
    G^{*}(t)&=&\langle\Psi(t)| e^{-iH_0t} |\Psi(t)\rangle,\nonumber\\
    &=&\cos^2\left(g_0\right) + \sin^2\left(g_0\right)\langle\sigma^z_1(t)\sigma^z_1(0) \rangle_0,
    \label{charac_func}
\end{eqnarray}
where $\sigma^z_1(t)=e^{iH_0t}\sigma^z_1e^{-iH_0t}$, $\langle \cdots\rangle_0\equiv \langle 0|\cdots|0\rangle$ and $^*$ denotes the complex conjugate. The first term in the above equation, which comes from the vacuum state contribution in Eq. (\ref{state}), will produce a trivial delta peak at $w=0$ in $P(w)$. This term can be removed if we take $g_0=\pi/2$. 

The second term in Eq. (\ref{charac_func}) encodes the nontrivial dynamics generated by the quench. Thus, interestingly, the work distribution $P(w)$ for our protocol is \textit{entirely} determined only by knowing the two-point autocorrelation function at the edge $\langle\sigma^z_1(t)\sigma^z_1(0) \rangle_0$. This represents a great advantage of the $\delta(t)$-kick protocol considered here. 
In fact, from this autocorrelation we can directly obtain the response function, $\Phi(t)/2\equiv-\text{Im}[\langle\sigma^z_1(t)\sigma^z_1(0)\rangle_0]$, where $\text{Im}$ denotes the imaginary part. For a \textit{general} weak perturbation, Ref. \cite{Gen_LRT_PRL_2024} showed that $P(w)$ can be fully determined once $\Phi(t)$ is known (we corroborate this finding in App. \ref{Appen1}).

An exact expression for $G(t)$ can be obtained once we express $\langle\sigma^z_1(t)\sigma^z_1(0) \rangle_0$ in terms of the fermionic modes that diagonalize $H_0$ [see Eq. (\ref{H_diag})] \cite{Moa_PRR_2025}. The characteristic function (\ref{charac_func}) becomes $G(t)=G_{\delta}+4\sin^2\left(g_0\right)\mc G(t)$, where $G_{\delta}=\cos^2\left(g_0\right) + \sin^2\left(g_0\right)\langle \sigma^z_1 \rangle^2_0$ is a time-independent term, and
\begin{eqnarray}
    \mc G(t)&=&\sum_{\kappa,\kappa'}f_{\kappa\kappa'}e^{i\left(\Gamma_\kappa + \Gamma_{\kappa'}\right)t},
    \label{charac_func2}
\end{eqnarray}
where $f_{\kappa\kappa'}=v_{\kappa}u_{\kappa'}(v_{\kappa}u_{\kappa'}-v_{\kappa'}u_{\kappa})$,
with $u_{\kappa}=(\phi_{1\kappa}+\psi_{1\kappa})/2$ and $v_{\kappa}=(\phi_{1\kappa}-\psi_{1\kappa})/2$. The amplitude $f_{\kappa\kappa'}$ satisfies $\sum_{\kappa,\kappa'}f_{\kappa\kappa'}=(1-\langle \sigma^z_1 \rangle^2_0)/4$, which directly gives us the normalization condition $G(0)=1$. Note that $G(t)$ gained another constant term $\propto\langle \sigma^z_1 \rangle^2_0$ that will contribute to the delta peak in $P(0)$. This term is responsible for describing a completely polarized chain, $\langle \sigma^z_1 \rangle_0|_{h\to\infty}\to 1$, and appears because in this situation we essentially have $[H_0|_{h\to\infty},\sigma^z_1]=0$ thus $|\Psi_0\rangle$ is also an eigenstate of $\sigma^z_1$. Far from this extreme limit, that is, for finite $h$, Eq. (\ref{charac_func2}) dominates over these other contributions. 

\begin{figure}
    \centering
    \includegraphics[scale=0.52]{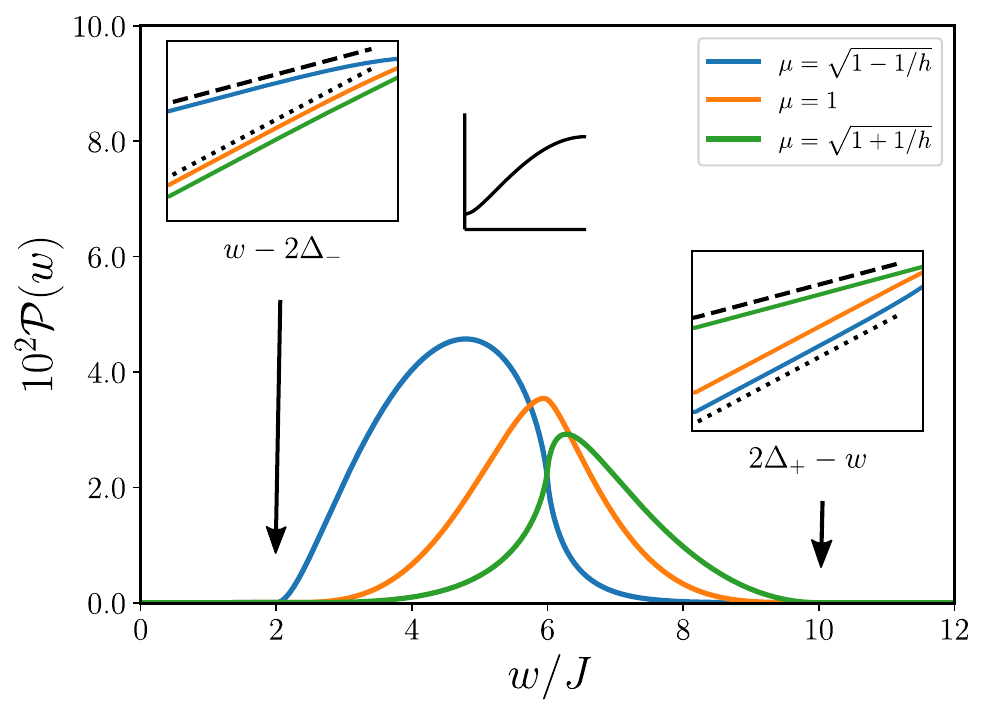}
    \caption{$\mc P(w)$ in the region $\sqrt{1-1/h}\leq\mu\leq\sqrt{1+1/h}$ with $h=1.5$. Insets highlight the low- and high-energy edge singularities of $\mc P(w)$, where the dashed and dotted black lines correspond to Eqs. (\ref{edge_sing_deloc_deloc}) and (\ref{edge_sing_boundary_h_right}) (see also the text below this equation). The system spectrum is indicated in the middle of the figure.}
    \label{fig_WorkDist_h1.5_Deloc}
\end{figure}

The inverse Fourier transform of Eq. (\ref{charac_func2}) can be taken as straightforward, and we conclude that the work distribution for our protocol has the general form
\begin{eqnarray}
    P(w)&=&G_{\delta}\delta(w) + \mc P(w),
    \label{work_statis_gen}
\end{eqnarray}
where the second term reads
\begin{eqnarray}
    \mc P(w)&=&4\sin^2\left(g_0\right)\sum_{\kappa,\kappa'}f_{\kappa\kappa'}\delta\left(w-\left(\Gamma_\kappa + \Gamma_{\kappa'}\right)\right).
    \label{work_sta}
\end{eqnarray}
The above equation clarifies the previously mentioned fact that the protocol (\ref{protocol}) creates two excitations. Interestingly, the protocol intensity $g_0$ appears in Eq. (\ref{work_sta}) only as an overall constant, and so the above expression remains valid even far from the perturbative regime $g_0\ll J$. Finally, notice the additive structure of Eq. (\ref{work_sta}), where the contribution arising purely from the bulk modes ($\kappa=k$ and $\kappa'=k'$) can be separated from the others. This allows us to better identify the edge state contributions to $\mc P(w)$.

In the next section, we show results obtained from the numerical evaluation of the summations in Eq. (\ref{work_sta}). As already mentioned, we assume a semi-infinite chain, which allows us to replace $\sum_{k}Q_k\to \frac{N}{\pi}\int_0^{\pi}dk\:Q(k)$ for the bulk modes. We also set $g_0=J$.

\section{Results}
\label{sec_res}

In this section, we provide the results for $\mc P(w)$ in Eq. (\ref{work_sta}) for the different boundary phases.

\subsection{\texorpdfstring{$h>1$ and $\sqrt{1-1/h}\leq\mu\leq\sqrt{1+1/h}$}{h > 1 and sqrt{1-1/h} leq mu leq sqrt{1+1/h}}}

\label{region1}

For this parameter range, the system only admits the bulk modes. For a semi-infinite chain $N\to \infty$, these modes build up an energy band $\Gamma_k=2\sqrt{1+h^2-2h\cos k}$ with $k\in[0,\pi]$ \cite{PRB_main_ref_2016}. The band gap is given by $\Gamma_{k=0}=2|h-1|\equiv \Delta_-$, while the highest energy is given by $\Gamma_{k=\pi}=2(h+1)\equiv \Delta_+$.

\begin{figure}
    \centering
    \includegraphics[scale=0.52]{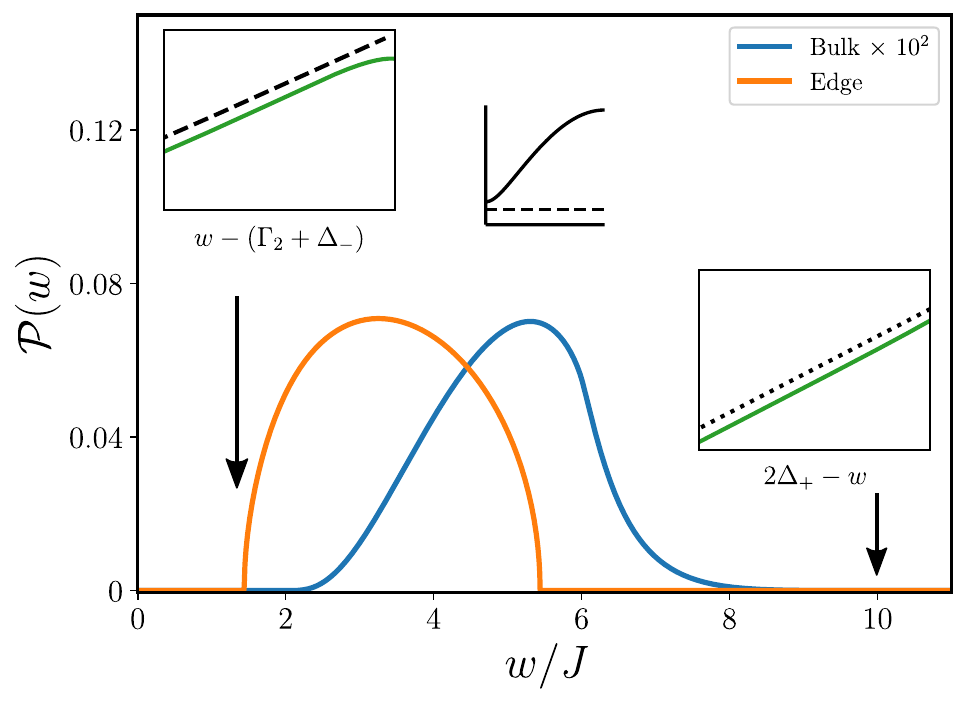}
    \caption{$\mc P(w)$ for $\mu=0.2$ and $h=1.5$ (in the region $\mu<\sqrt{1-1/h}$ and $h>1$). The blue line represents the pure bulk states contribution (times $10^2$), while the orange one is the edge state contribution (their sum gives $\mc P(w)$; see the green line in the insets). The insets, plotted on a log-log scale, highlight the low- and high-energy edge singularities (ESs) of $\mc P(w)$, where the dashed and dotted black lines correspond to Eq.~(\ref{edge_sing_mode2}). The system spectrum is indicated in the middle of the figure, where the dashed line marks the energy of the edge state $\gamma_2$.}
    \label{fig_WorkDist_h1.5_mu_mode2}
\end{figure}

In Fig. \ref{fig_WorkDist_h1.5_Deloc} the results for $\mc P(w)$ are shown. Firstly, in the absence of the impurity ($\mu=1$), the work distribution is symmetric around $w/J = 6$. As $\mu$ deviates from 1, $\mc P(w)$ becomes skewed towards the low- or high-energy region, depending on whether $\mu < 1$ or $\mu > 1$, respectively. Exactly at the boundaries $\mu = \sqrt{1\pm 1/h}$, this effect becomes more pronounced. Importantly, this reshaping of $\mc P(w)$ arises solely from the wave functions $\phi_{1k}$ and $\psi_{1k}$, since the system spectrum remains unchanged in the considered region (see Fig. \ref{fig_WorkDist_h1.5_Deloc}).

Although the exact functional form of $\mc P(w)$ for arbitrary $w$ and $\mu$ is difficult to obtain, in the regions near the low-energy threshold $w\approx2\Delta_-$ and the high-energy limit $w\approx2\Delta_+$, $\mc P(w)$ shows universal behavior (in the sense that it does not depend on $\mu$ continuously). The insets of Fig. \ref{fig_WorkDist_h1.5_Deloc} highlight $\mc P(w)$ in these regions. The log-log graphs reveal the characteristic edge singularities (ESs) that appear at zero temperature \cite{gambassi2011}, where $\mc P(w)$ goes to zero as a power-law with a given exponent \cite{PRL_Ale_2008,PRL_Ale_2012,PRE_Ale_2013,PRE_Ale_2013_2,PRB_Dora_2024}. Here, we see that the exponent only depends on whether $\mu$ is at the boundaries $\mu=\sqrt{1\pm1/h}$ or far from it. 

We can determine the exact exponents of the ESs by analyzing the temporal decay of the characteristic function $\mc G(t)$ \cite{gambassi2011,Russomanno_2016}. Using the stationary phase approximation (SPA) \cite{PRB_Dutta_2022}, we find that in next-leading order $\mc G(|t|\gg J^{-1})\sim i\sgn(t)\left(e^{2i\Delta_- t} - e^{2i\Delta_+ t}\right)/|t|^5$, for $\sqrt{1-1/h}<\mu<\sqrt{1+1/h}$ and $h>1$, where $\sgn(t)$ is the signal function. Taking the inverse Fourier transform of this result, we obtain
\begin{equation}
    \mc P(w)\sim\begin{cases} 
          (w-2\Delta_-)^4\:\Theta\left(w-2\Delta_-\right), & w\approx 2\Delta_- \\
          (2\Delta_+-w)^4\:\Theta\left(2\Delta_+-w\right), & w\approx2\Delta_+
       \end{cases}
       \label{edge_sing_deloc_deloc}
\end{equation}
where $\Theta(x)$ is the Heaviside step function. This exponent $4$ is a consequence of the quadratic band dispersion $\Gamma_k$ around $k=0,\pi$ and that $f_{kk'}|_{(0,0)}\sim k^4k'^4$ (similarly around $\pi$). The dotted black lines in the insets of Fig. \ref{fig_WorkDist_h1.5_Deloc} confirm Eq. (\ref{edge_sing_deloc_deloc}).

\begin{figure}
    \centering
    \includegraphics[scale=0.52]{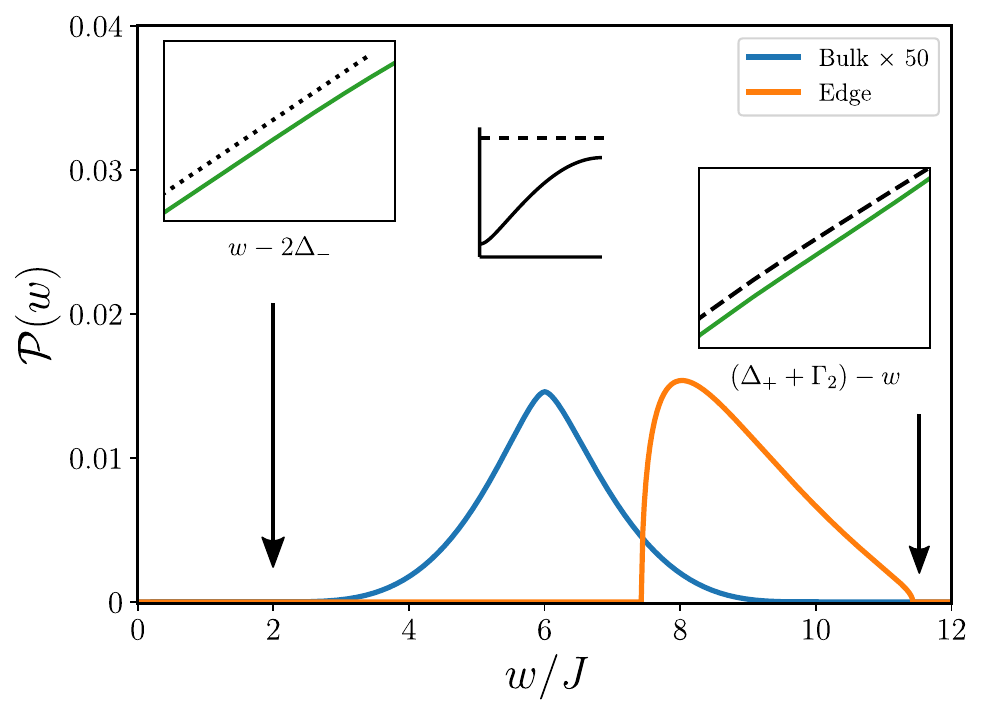}
    \caption{$\mc P(w)$ for $\mu = 2.0$ and $h = 1.5$ (in the region $\mu > \sqrt{1 + 1/h}$ and $h > 1$). The same color scheme as in Fig.~\ref{fig_WorkDist_h1.5_mu_mode2} is used. The dashed and dotted black lines in the insets correspond to Eq.~(\ref{edge_sing_mode2_2}). The system spectrum is shown in the middle of the figure, with the dashed line indicating the energy of the edge state $\gamma_2$.} 
    \label{fig_WorkDist_h1.5_mu_mode2_2}
\end{figure}

In a similar way, at the boundary $\mu=\sqrt{1-1/h}$ with $h>1$, we have $\mc G(|t|\gg J^{-1})\sim -i\sgn(t)\left(e^{2i\Delta_- t}/|t|^3 + e^{2i\Delta_+ t}/|t|^5\right)$, which yields
\begin{equation}
    \mc P(w)\sim\begin{cases} 
          (w-2\Delta_-)^2\:\Theta\left(w-2\Delta_-\right), & w\approx 2\Delta_- \\
          (2\Delta_+-w)^4\:\Theta\left(2\Delta_+-w\right), & w\approx2\Delta_+.
       \end{cases}
       \label{edge_sing_boundary_h_right}
\end{equation}
We see that the ES in the low-energy threshold is altered, exhibiting now an exponent $2$, whereas that in the high-energy region remains unaffected. This is attributed to the edge mode $\gamma_2$, which for $\mu < \sqrt{1 - 1/h}$ emerges with energy $\Gamma_2 < \Delta_-$. For the second boundary, $\mu=\sqrt{1+1/h}$ with $h>1$, we obtain $\mc P(w)\sim(w-2\Delta_-)^4\Theta(w-2\Delta_-)$ for $w\approx 2\Delta_-$ and $\mc P(w)\sim (2\Delta_+-w)^2\Theta(2\Delta_+-w)$ for $w\approx 2\Delta_+$. This time is the high-energy region that is altered, due to $\gamma_2$ emerging with energy $\Gamma_2 > \Delta_+$ for $\mu > \sqrt{1+1/h}$. Again, it is worth stressing that these behaviors at the boundaries come purely from the modifications of the wave functions at the edge. The dashed and dotted black lines in the insets of Fig. \ref{fig_WorkDist_h1.5_Deloc} show the agreement between the numerical results and the above analytical predictions.

\subsection{\texorpdfstring{$h>1$ and $\mu\lessgtr\sqrt{1\mp1/h}$}{h > 1 and mu lessgtr sqrt(1 + 1/h)}}


Now, let us move to the situation where the system spectrum changes due to the appearance of the edge mode $\gamma_2$.

For the parameter range $h>1$ and $\mu<\sqrt{1-1/h}$, as mentioned above, $\gamma_2$ emerges with energy $\Gamma_{2}<\Delta_-$ \cite{PRB_main_ref_2016}. Thus, we can already expect that the low-energy region of $\mc P(w)$ is the one more affected.  From the results shown in Fig. \ref{fig_WorkDist_h1.5_mu_mode2} we, in fact, confirm this.

\begin{figure}
    \centering
    \includegraphics[scale=0.52]{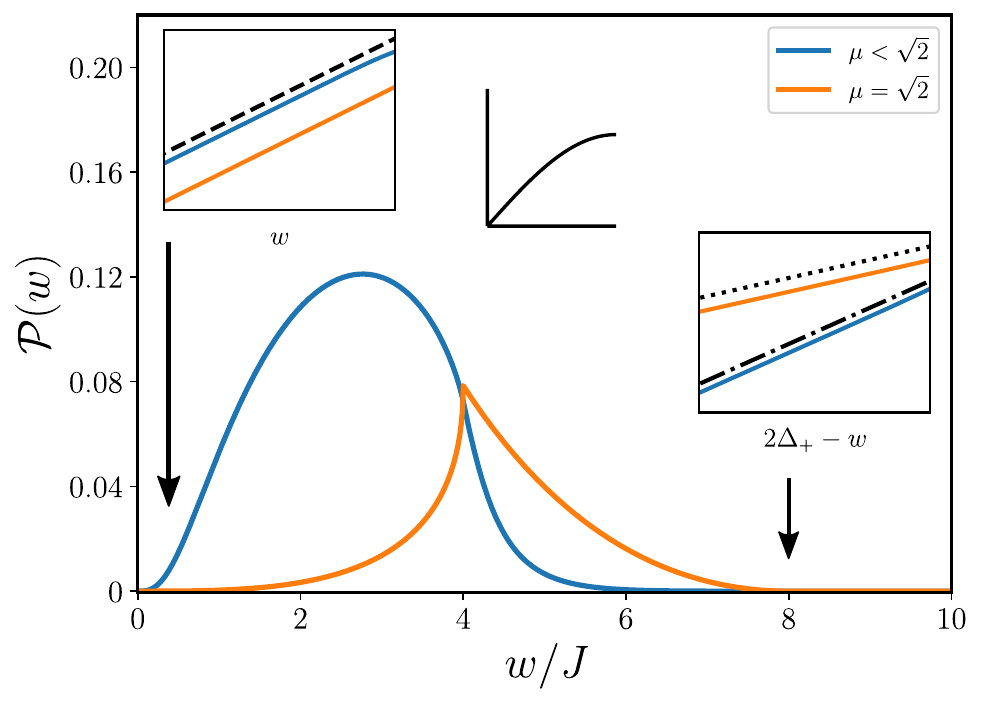}
    \caption{$\mc P(w)$ in the region $\mu\leq \sqrt{1+1/h}=\sqrt{2}$ with $h=1$. For the blue curve, $\mu=0.5$. The dashed, dotted and dashed-dotted black lines in the insets correspond to Eq.~(\ref{edge_h_1_1}) and Eq.~(\ref{edge_h_1_2}). The system spectrum (gapless) is indicated in the middle of the figure.}
    \label{fig_WorkDist_critical}
\end{figure}

The ESs again follow from the behavior of $\mc G(t)$ for $t\gg J^{-1}$. Using SPA, we obtain $\mc G(|t|\gg J^{-1})\sim e^{-i\sgn(t)\pi/4}e^{i(\Delta_-+\Gamma_{2})t}/|t|^{3/2}-i\sgn(t)e^{2i\Delta_+ t}/|t|^5$. Notice the slower $t^{-3/2}$ decay in $\mc G(t)$. Thus, we now have
\begin{equation}
    \mc P(w)\sim\begin{cases} 
          \sqrt{w-\Gamma_{2}-\Delta_-}\:\Theta\left(w-\Gamma_{2}-\Delta_-\right),\\
          (2\Delta_+-w)^4\:\Theta\left(2\Delta_+-w\right),
       \end{cases}
       \label{edge_sing_mode2}
\end{equation}
where the first line is valid only when $w\approx \Gamma_{2}+\Delta_-$ while the second when $w\approx2\Delta_+$. As expected, the ES in high-energy remains the same, but near the low-energy threshold we see a new $1/2$ exponent. This new exponent comes from the $t^{-3/2}$ term in $\mc G(t)$. The dashed and dotted black lines in the insets of Fig. \ref{fig_WorkDist_h1.5_mu_mode2} confirm the above result.

For the parameter range $h>1$ and $\mu>\sqrt{1+1/h}$, $\gamma_2$ has energy $\Gamma_{2}>\Delta_+$ \cite{PRB_main_ref_2016}. Similarly, the asymptotic behavior of $\mc P(w)$ reads
\begin{equation}
    \mc P(w)\sim\begin{cases} 
          (w-2\Delta_-)^4\:\Theta\left(w-2\Delta_-\right),\\
          \sqrt{\Gamma_{2}+\Delta_+-w}\:\Theta\left(\Gamma_{2}+\Delta_+-w\right),
       \end{cases}
       \label{edge_sing_mode2_2}
\end{equation}
for $w\approx2\Delta_-$ and $w\approx\Gamma_{2}+\Delta_+$, respectively. We see again that the edge mode $\gamma_2$  modifies the exponent of the ES [compare to Eq. (\ref{edge_sing_deloc_deloc})], this time in the high-energy region. In Fig. \ref{fig_WorkDist_h1.5_mu_mode2_2} we show  the agreement between the numerical and analytical results.

\begin{figure}
    \centering
    \includegraphics[scale=0.52]{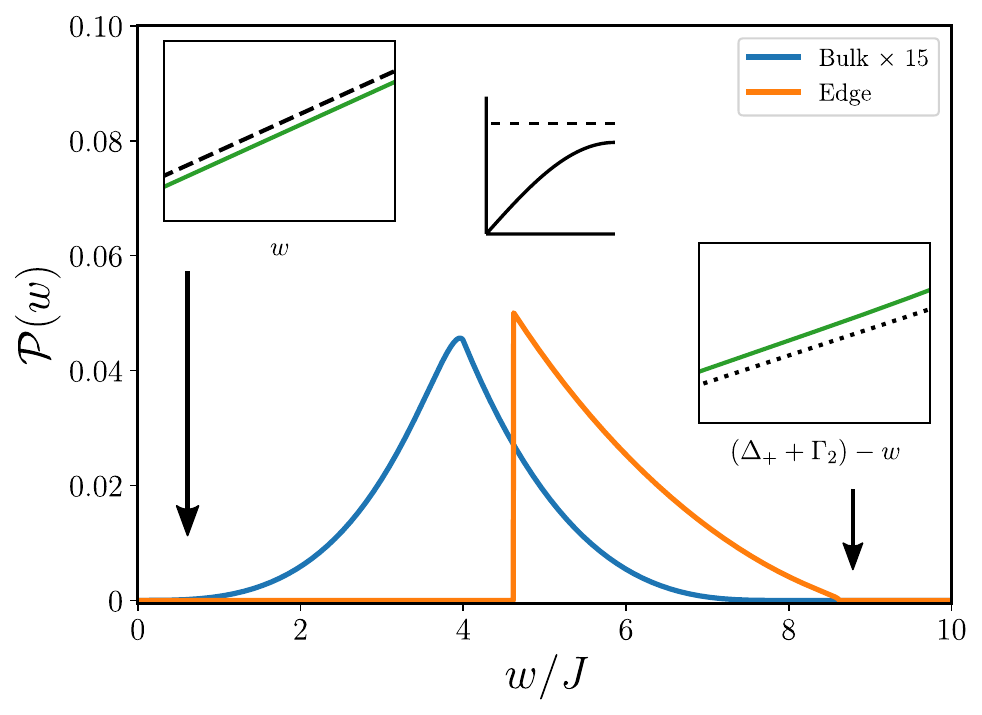}
    \caption{$\mc P(w)$ for $\mu=2$ and $h=1$ (in the region $\mu> \sqrt{1+1/h}=\sqrt{2}$). The same color scheme as in Fig.~\ref{fig_WorkDist_h1.5_mu_mode2} is used. The dashed and dotted black lines in the insets correspond to Eq.~(\ref{edge_h_1_3}). The system spectrum (gapless) is indicated in the middle of the figure, where the dashed line marks the energy of the edge state $\gamma_2$.}
    \label{fig_WorkDist_critical_2}
\end{figure}

The results in Eqs. (\ref{edge_sing_mode2}) and (\ref{edge_sing_mode2_2}) show how the ES of the square-root type is a clear signature of the edge state, whatever it appears inside the band gap ($\Gamma_2<\Delta_-$) or above the bulk band ($\Gamma_2>\Delta_+$).

\subsection{\texorpdfstring{$h=1$ and $\mu<\sqrt{2}$}{h = 1 and mu < sqrt(2)}}

We now consider the critical line, $h=1$, where the gapless bulk spectrum is expected to quantitatively modify the low-energy behavior of $\mc P(w)$. For $\mu<\sqrt{1+1/h}=\sqrt{2}$, the system does not support an edge mode, and the work distribution is entirely governed by the bulk modes.  

Using SPA, we find the long-time asymptotic behavior $\mc G(|t|\gg J^{-1})\sim 1/t^4-i\sgn(t)e^{2i\Delta_+ t}/|t|^5$. This result leads to
\begin{equation}
    \mc P(w)\sim\begin{cases} 
          w^3\:\Theta(w),& w\approx0\\
          (2\Delta_+-w)^4\:\Theta\left(2\Delta_+-w\right), & w\approx2\Delta_+.
       \end{cases}
       \label{edge_h_1_1}
\end{equation}
Therefore, at criticality the low-energy ES is characterized by the exponent $3$, instead of the exponent $4$ obtained in the gapped phase [see Eq. (\ref{edge_sing_deloc_deloc})]. By contrast, the high-energy ES retains the exponent $4$, reflecting the fact that $\Gamma_k$ remains quadratic in the vicinity of $k=\pi$. The numerical results presented in Fig. \ref{fig_WorkDist_critical} are fully consistent with these analytical predictions. 

The scaling $\mc P(w\ll J)\sim w^3$ follows from the $1/t^4$ contribution to $\mc G(t)$, which in turn originates from $f_{kk'}|_{(0,0)}\sim k'(k'-k)$. The same result can be understood within the Ising field theory \cite{sachdev1999quantum}. Open boundary conditions imply that, in the continuum limit, $\sigma^z_1\sim \chi(0)\partial_x\chi(0)$, where $\chi(x)$ denotes a chiral massless Majorana field \cite{Diehl_PRB_1983,CARDY1984514,OSHIKAWA1997533}. Since $\chi(x)$ has scaling dimension $1/2$ \cite{sachdev1999quantum}, the boundary operator $\chi(0)\partial_x\chi(0)$ has scaling dimension $2$, implying the asymptotic decay $\langle\sigma^z_1(t)\sigma^z_1(0)\rangle_0\sim t^{-4}$.

\begin{figure}
    \centering
    \includegraphics[scale=0.52]{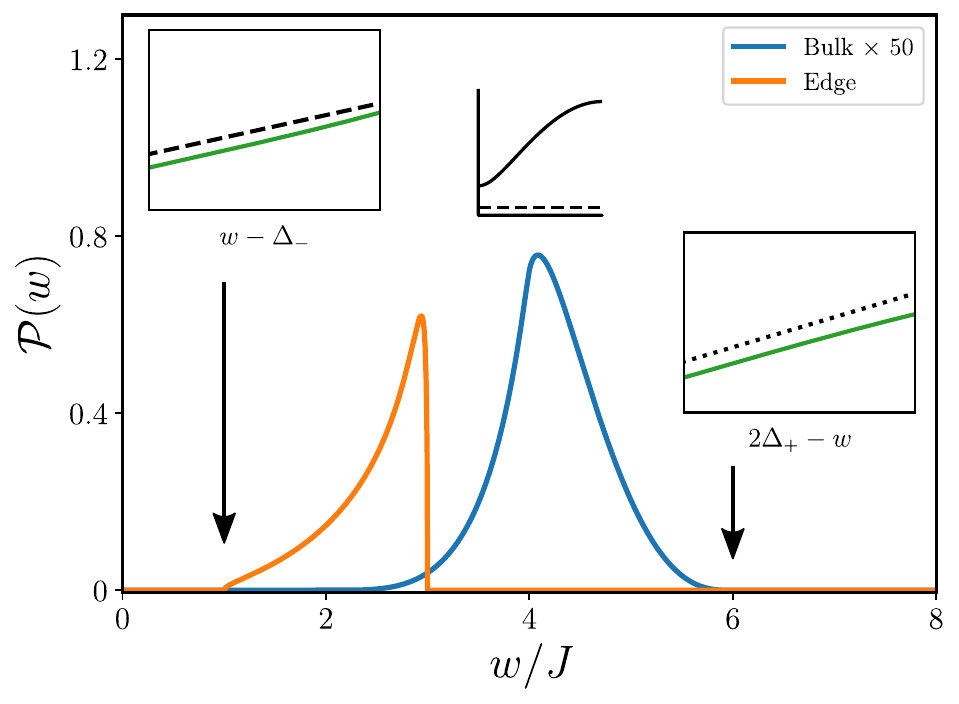}
    \caption{$\mc P(w)$ for $\mu=1.5$ and $h=0.5$ (in the region $\mu<\sqrt{1+1/h}$ and $h<1$). The same color scheme as in Fig.~\ref{fig_WorkDist_h1.5_mu_mode2} is used. The dashed and dotted black lines in the insets correspond to Eq.~(\ref{edge_sing_mode1}). The system spectrum is indicated in the middle of the figure, where the dashed line marks the exponentially small energy of the edge state $\gamma_1$ in this case.}
    \label{fig_WorkDist_h0.5_mu_mode1}
\end{figure}

\subsection{\texorpdfstring{$h=1$ and $\mu=\sqrt{2}$}{h = 1 and mu = sqrt(2)}}

As in the previous case, the work distribution is still entirely determined by the bulk modes. However, the edge mode that emerges for $\mu>\sqrt{2}$ is expected to modify the high-energy ES, similarly to the case $\mu=\sqrt{1+1/h}$ with $h>1$ [see the discussion below Eq. (\ref{edge_sing_boundary_h_right})]. 

The SPA now yields, $\mc G(|t|\gg J^{-1})\sim 1/t^4+e^{i\sgn(t)\pi/4}e^{i\Delta_+t}/|t|^{3/2}-i\sgn(t)e^{2i\Delta_+ t}/|t|^3$. This immediately gives
\begin{equation}
    \mc P(w)\sim\begin{cases} 
          w^3\:\Theta(w),& w\approx0\\
          (2\Delta_+-w)^2\:\Theta\left(2\Delta_+-w\right), & w\approx2\Delta_+.
       \end{cases}
       \label{edge_h_1_2}
\end{equation}
As anticipated, the high-energy ES is modified, while in low-energy we still see the $w^3$ scaling. Fig. \ref{fig_WorkDist_critical} displays $\mc P(w)$ together with the corresponding ESs. Interestingly, the derivative of $\mc P(w)$ develops a singularity at the mid-energy $w=\Delta_+$. More precisely, one finds $\mc P'(w)\sim(\Delta_+-w)^{-1/2}$ in the vicinity of $w=\Delta_+$.

\subsection{\texorpdfstring{$h=1$ and $\mu>\sqrt{2}$}{h = 1 and mu > sqrt(2)}}

In this parameter regime, the edge mode $\gamma_2$ emerges with energy $\Gamma_2>\Delta_+$. Its presence is therefore expected to modify the high-energy sector of the work distribution, similarly to the case $\mu>\sqrt{1+1/h}$ with $h>1$ (see Fig. \ref{fig_WorkDist_h1.5_mu_mode2_2}).

The corresponding numerical results are shown in Fig. \ref{fig_WorkDist_critical_2}. In comparison with Fig. \ref{fig_WorkDist_critical}, the impurity produces a noticeable reshaping of the bulk contribution. The ESs are determined by the long-time asymptotic behavior $\mc G(|t|\gg J^{-1})\sim 1/t^4-e^{-i\sgn(t)\pi/4}e^{i(\Gamma_2+\Delta_+)t}/|t|^{3/2}$, which implies
\begin{equation}
    \mc P(w)\sim\begin{cases} 
          w^3\:\Theta(w),\\
          \sqrt{\Gamma_2+\Delta_+-w}\:\Theta\left(\Gamma_2+\Delta_+-w\right),
       \end{cases}
       \label{edge_h_1_3}
\end{equation}
where the first line is valid only when $w\approx0$ while the second when $w\approx\Gamma_2+\Delta_+$. Therefore, the high-energy sector exhibits the characteristic square-root-type ES associated with the edge mode.

\subsection{\texorpdfstring{$h<1$ and $\mu<\sqrt{1+1/h}$}{h < 1 and mu < sqrt(1 + 1/h)}}


For this case, the edge mode $\gamma_1$ appears as a zero-energy mode (assuming a semi-infinite chain), $\Gamma_1=0$.

The results for this case are shown in Fig. \ref{fig_WorkDist_h0.5_mu_mode1}. Once again, we see that the impurity strength $\mu$ strongly reshapes $\mc P(w)$. This is more pronounced in the low-energy region, given the presence of the edge state.

\begin{figure}
    \centering
    \includegraphics[scale=0.52]{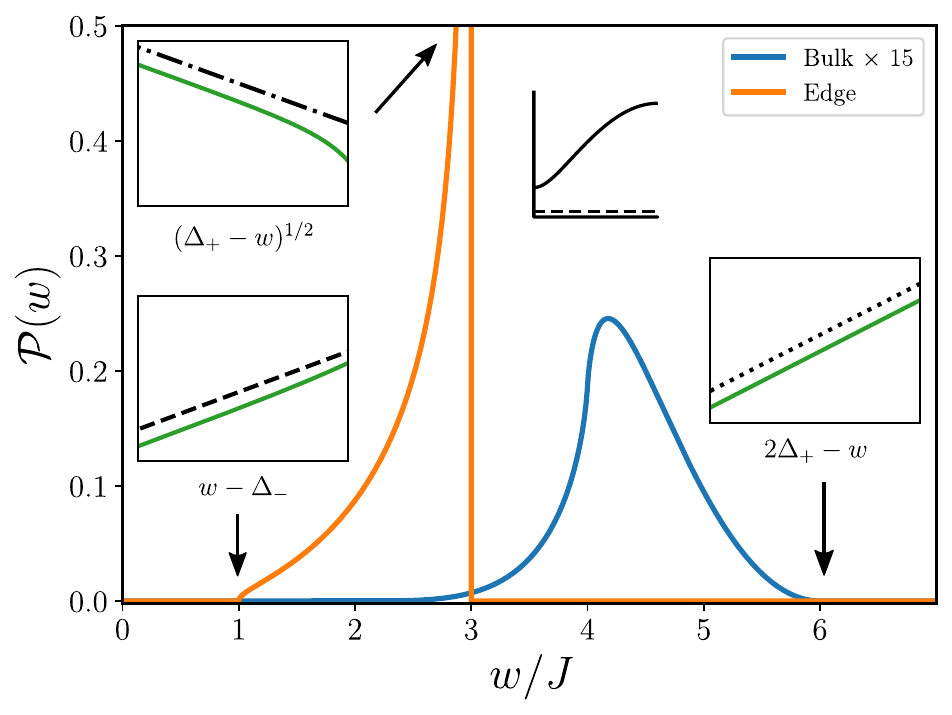}
    \caption{$\mc P(w)$ for $\mu=\sqrt{1+1/h}$ and $h=0.5$. The same color scheme as in Fig.~\ref{fig_WorkDist_h1.5_mu_mode2} is used. The dashed, dotted and dashed-dotted black lines in the insets correspond to Eq.~(\ref{edge_sing_mode1_boundary}). Again, the system spectrum is indicated in the middle of the figure.}
    \label{fig_WorkDist_h0.5_mu_mode1_Bound}
\end{figure}

In order to determine the behavior of $\mc P(w)$ near the ESs, we perform the same SPA analysis. In this case, we obtain $\mc G(|t|\gg J^{-1})\sim e^{-i\sgn(t)\pi/4}e^{i\Delta_-t}/|t|^{3/2}-i\sgn(t)e^{2i\Delta_+ t}/|t|^5$, which leads us to the result
\begin{equation}
    \mc P(w)\sim\begin{cases} 
          \sqrt{w-\Delta_-}\:\Theta\left(w-\Delta_-\right), & w\approx \Delta_- \\
          (2\Delta_+-w)^4\:\Theta\left(2\Delta_+-w\right), & w\approx2\Delta_+.
       \end{cases}
       \label{edge_sing_mode1}
\end{equation}
Similarly, near the low-energy threshold we obtain the same exponent $1/2$ while the high-energy region remains dominate by the bulk modes [see Eq. (\ref{edge_sing_mode2})]. Therefore, although the modes $\gamma_1$ and $\gamma_2$ have very distinct physical origins, this is not reflected in the ESs of $\mc P(w)$. But the general behavior of $\mc P(w)$, for all energies scales, does change.

\subsection{\texorpdfstring{$h<1$ and $\mu=\sqrt{1+1/h}$}{h < 1 and mu = sqrt(1 + 1/h)}}


Here, when compared to the previous case, the system spectrum does not change, we still have a gapped bulk band, $\Gamma_k$, and an edge state with zero energy, $\Gamma_1 = 0$. However, the wave functions at the edge, $j = 1$, are modified due to the emergence of $\gamma_2$ for $\mu > \sqrt{1 + 1/h}$. Thus, similarly to what happened at the boundaries $\mu = \sqrt{1 \mp 1/h}$, we can expect that $\mc P(w)$ is also nontrivially affected.

The behavior of $\mc{P}(w)$ is shown in Fig. \ref{fig_WorkDist_h0.5_mu_mode1_Bound}. Interestingly, in addition to the ESs in the low- and high-energy regions, we observe that at mid-energy, $w = \Delta_{+} + \Gamma_1$, $\mc P(w)$ develops a sharp singularity. This corresponds to the creation of one excitation with energy $\Gamma_1=0$ and another with energy $\Gamma_{k=\pi} = \Delta_{+}$ at the top of the bulk band [see Eq. (\ref{work_sta})].

\begin{figure}
    \centering
    \includegraphics[scale=0.52]{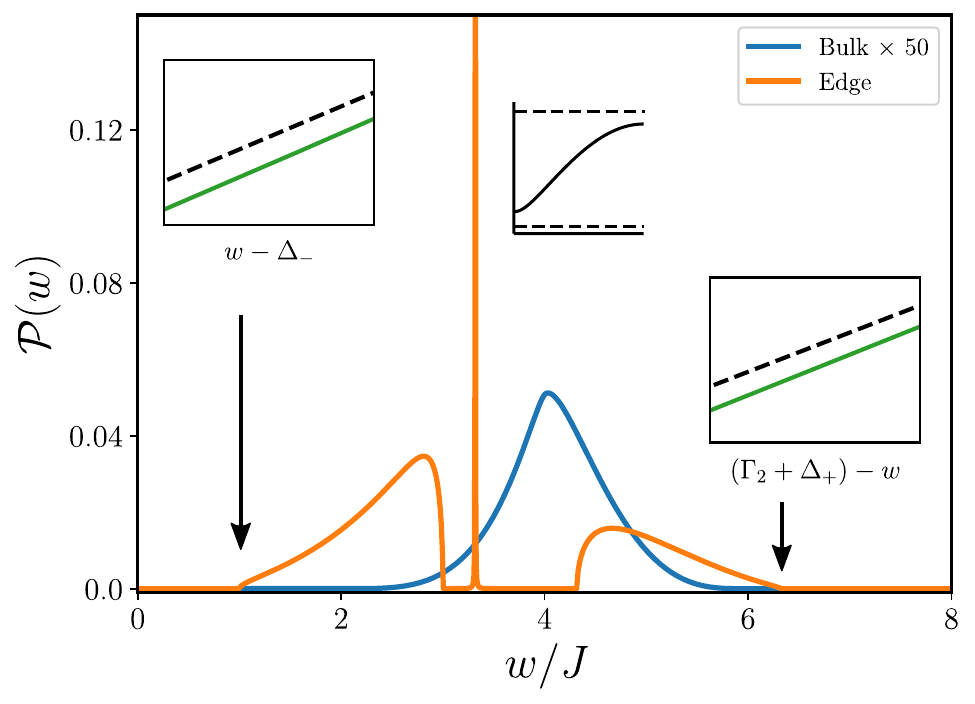}
    \caption{$\mc P(w)$ for $\mu=2.5$ and $h=0.5$ (in the region $\mu>\sqrt{1+1/h}$ and $h<1$). The same color scheme as in Fig.~\ref{fig_WorkDist_h1.5_mu_mode2} is used. The dashed black lines in the insets correspond to Eq.~(\ref{edge_sing_mode1_2}). The system spectrum, with the two edge states, is indicated in the middle of the figure.}
    \label{fig_WorkDist_h0.5_mu_mode1_2}
\end{figure}

\begin{table*}
\scalebox{1.0}{
\begin{tabular}{|c|c|c|c|}
	\hline
	Region & Low-energy &Mid-energy & High-energy \\
	\hline\hline
	$h>1$ and $\sqrt{1-1/h}<\mu<\sqrt{1+1/h}$ & $(w-2\Delta_-)^4$ & - & $(2\Delta_+-w)^4$ \\
	\hline
	$h>1$ and $\mu=\sqrt{1-1/h}$ & $(w-2\Delta_-)^2$ & - & $(2\Delta_+-w)^4$ \\
	\hline
	$h>1$ and $\mu=\sqrt{1+1/h}$ & $(w-2\Delta_-)^4$ & - & $(2\Delta_+-w)^2$ \\
	\hline
	$h>1$ and $\mu<\sqrt{1-1/h}$ & $\sqrt{w-\Gamma_2-\Delta_-}$ & - & $(2\Delta_+-w)^4$ \\
	\hline
    $h>1$ and $\mu>\sqrt{1+1/h}$ & $(w-2\Delta_-)^4$ & - & $\sqrt{\Gamma_2+\Delta_+-w}$ \\
	\hline
    $h=1$ and $\mu<\sqrt{2}$ & $w^3$ & - & $(2\Delta_+-w)^4$ \\
    \hline
    $h=1$ and $\mu=\sqrt{2}$ & $w^3$ & - & $(2\Delta_+-w)^2$ \\
    \hline
    $h=1$ and $\mu>\sqrt{2}$ & $w^3$ & - & $\sqrt{\Gamma_2+\Delta_+-w}$ \\
	\hline
    $h<1$ and $\mu<\sqrt{1+1/h}$ & $\sqrt{w-\Delta_-}$ & - & $(2\Delta_+-w)^4$ \\
    \hline
	$h<1$ and $\mu=\sqrt{1+1/h}$ & $\sqrt{w-\Delta_-}$ & $\left(\Delta_+-w\right)^{-1/2}$ & $(2\Delta_+-w)^2$ \\
	\hline
	$h<1$ and $\mu>\sqrt{1+1/h}$ & $\sqrt{w-\Delta_-}$  & $\delta(w-\Gamma_2)$ & $\sqrt{\Gamma_2+\Delta_+-w}$ \\
	\hline
\end{tabular}}
\caption{Edge singularities (ESs) and mid-energy singularities of the quantum work distribution (\ref{work_sta}) for the different boundary phases of the model in Eq.~(\ref{main_Hamil}).}
\label{t1}
\end{table*}

The exact form of this mid-energy singularity can also be obtained from SPA. For this case, we obtain that $\mc G(|t|\gg J^{-1})\sim e^{-i\sgn(t)\pi/4}\left[e^{i\Delta_-t}/|t|^{3/2}+e^{i\Delta_+t}/\sqrt{|t|}\right]-i\sgn(t)e^{2i\Delta_+ t}/|t|^3$. Thus,

\begin{equation}
    \mc P(w)\sim\begin{cases} 
          \sqrt{w-\Delta_-}\:\Theta\left(w-\Delta_-\right), & w\approx \Delta_- \\
          \frac{1}{\sqrt{\Delta_+ -\: w}}\:\Theta\left(\Delta_+-w\right), & w\approx \Delta_+ \\
          (2\Delta_+-w)^2\:\Theta\left(2\Delta_+-w\right), & w\approx2\Delta_+.
       \end{cases}
       \label{edge_sing_mode1_boundary}
\end{equation}
We see that the mid-energy singularity has the form $(\Delta_+ -w)^{-1/2}$, while the ESs near the low- and high-energy regions have exponent $1/2$ and $2$, respectively. The slower $\sqrt{t}$ term in $\mc G(t)$ is responsible for the appearance of the mid-energy singularity.

\subsection{\texorpdfstring{$h<1$ and $\mu>\sqrt{1+1/h}$}{h < 1 and mu > sqrt(1 + 1/h)}}


Finally, in this last parameter range, both edge modes $\gamma_{1,2}$ appear. This situation, unlike all cases discussed so far, makes $\mc P(w)$ acquire a contribution that comes purely from the two edge states (remember that $\mc P(w)$ accounts for pairs of excitations).

Fig. \ref{fig_WorkDist_h0.5_mu_mode1_2} shows the results. We see the familiar ESs in the low- and high-energy regions. However, this time, both have the same exponent $1/2$ due to the presence of the two edge states (one below and another above the bulk band). This becomes clear from the SPA result, which gives $\mc G(|t|\gg J^{-1})\sim e^{-i\sgn(t)\pi/4}e^{i\Delta_- t}/|t|^{3/2} + e^{i\Gamma_2 t} -e^{i\sgn(t)\pi/4}e^{i(\Delta_+ + \Gamma_2)t}/|t|^{3/2}$. Then,
\begin{equation}
    \mc P(w)\sim\begin{cases} 
          \sqrt{w-\Delta_-}\:\Theta\left(w-\Delta_-\right),  \\
          \delta(w-\Gamma_2),  \\
          \sqrt{\Gamma_2+\Delta_+-w}\:\Theta\left(\Gamma_2+\Delta_+-w\right),
       \end{cases}
       \label{edge_sing_mode1_2}
\end{equation}
where the first line is valid only when $w \approx \Delta_-$, the second when $w\approx\Gamma_2$, and the third when $w \approx \Delta_+ + \Gamma_2$. The delta peak at energy $w = \Gamma_2$ reflects the above-mentioned fact that $\mc P(w)$ acquires a purely edge state contribution. Physically, it arises from creating one excitation with energy $\Gamma_1 = 0$ and another with energy $\Gamma_2 > \Delta_+$ [notice the term $e^{-i\Gamma_2 t}$ in $\mc G(t)$].

\section{Concluding remarks}
\label{sec_conclu}

This work aimed to explore nonequilibrium signatures of boundary phases. We considered an exactly solvable Ising-like spin-$1/2$ chain with open boundary conditions (OBC) and an edge impurity. The combination of OBC with the impurity led to a rich equilibrium boundary phase diagram featuring up to two edge states. By preparing the system in its ground state and then suddenly kicking the impurity with an external magnetic field, we numerically computed the associated quantum work statistics using a two-time measurement (TTM) scheme. We showed that the work distribution is strongly affected by the impurity and by the presence of edge states. We also obtained analytical results, allowing us to determine the low- and high-energy edge singularities (ESs) across the different phases (see Table \ref{t1}). The exponents of these ESs are robust against quench protocols of finite duration, as shown in App. \ref{Appen1}. Moreover, this work demonstrated that the presence of edge states leaves clear and characteristic fingerprints in the quantum work distribution.

Experimentally, the verification of our results requires platforms that satisfy the following conditions. First, the ability to prepare the ground state of the Hamiltonian (\ref{main_Hamil}) is essential. This, in turn, demands a high degree of control over the local fields applied at the lattice sites, since an impurity must be engineered near the edge. Second, the platform must allow for the generation of nonequilibrium dynamics via a quench. These two requirements are within the operational reach of present-day Rydberg quantum simulators \cite{Bernien2017,Keesling2019,Rydn_exp_2021,PRXQuantum.2.017003}. Finally, and more challengingly, the implementation of TTM protocol is required. However, for the $\delta(t)$-kick protocol---and even for quenches of finite duration (see App. \ref{Appen1})---the characteristic function is fully determined by the autocorrelation function $\langle\sigma^z_1(t)\sigma^z_1(0)\rangle_0$ [see Eq. (\ref{charac_func})], which can also be measured with Rydberg atoms \cite{Rydb_PRX_2017,w1cp-l5vq} or with trapped ions \cite{Dumitrescu2022}. Therefore, we expect that the predictions of the present work can be accessed with Rydberg quantum simulators.


In the future, a direction to pursue is investigating the Floquet version of the problem presented here. Instead of only one $\delta$-kick, we can consider periodic kicks, as was done, for example, in Refs. \cite{Russomanno_2016, PhysRevLett.124.155302}, and then determine $P(w)$. Another interesting possibility would be to analyze different boundary physics, as the one considered in Ref. \cite{t78z-pyfr}.

\acknowledgements
It is a pleasure to thank R. G. Pereira, S. Deffner, K. Zawadzki, M. V. S. Bonan\c{c}a, S. Campbell and E. Miranda for useful comments. This work was supported by Conselho Nacional de Desenvolvimento Cient\'{i}fico e Tecnol\'{o}gico (CNPq), Brazil, through grant No. 200267/2023-0 and the John Templeton Foundation under Grant No.
62422. The work was also financed in part by the S\~{a}o Paulo Research Foundation (FAPESP), Brazil, Process No. 2022/15453-0.

\appendix

\section{Finite-time quench}
\label{Appen1}

In this Appendix, we analyze the effects of a finite-time quench on the exponents of the edge singularities. 

Let us now consider that 
\begin{equation}
    \mc H(t) = H_0 + (-g_0)\lambda(t)\sigma^z_1.
    \label{protocol2}
\end{equation}
The protocol $\lambda(t)$ has the general form: $\lambda(t)=0$ for $t\leq0$, $\lambda(t)\neq 0$, for $0<t<\tau$ and $\lambda(t)=0$ for $t\geq \tau$. We assume that for $t\leq 0$ the system is in its ground-state. 

To determine the quantum work distribution, we use again the two-time projective energy measurement scheme. The first energy measurement is performed at time $t=0$, which returns the energy of the ground state of $H_0$. The second energy measurement is performed at time $t=\tau$, which returns one of the eigen-energies of $H_0$ with probability $|\langle \Psi_m|\Psi(\tau)\rangle|^2$, where $|\Psi_m\rangle$ are the many-body eigenstates of $H_0$. We construct the quantum work distribution through
\begin{equation}
P(w,\tau)=\sum_m|\langle\Psi_m|\Psi(\tau)\rangle|^2\delta\left(w-(E_m-E_0)\right).
\end{equation}
Note that, in this case, $P(w,\tau)$ is also a function of the protocol duration $\tau$. In what follows, we take $E_0=0$.

For weak perturbations, $g_0\ll J$, we can obtain a perturbative expression for the characteristic function $G(s,\tau)$. Defining $P(w,\tau)\equiv\int ds\: e^{-iws}G(s,\tau)$, that is, as the Fourier transform of $G(s,\tau)$, it is straightforward to obtain that $G(s,\tau)=\langle\Psi(\tau)| e^{iH_0s} |\Psi(\tau)\rangle$. By expanding $|\Psi(\tau)\rangle$ in powers of $g_0$ we will arrive at
\begin{eqnarray}
    G(s,\tau)&\simeq& 1 -g^2_0\int_0^\tau dtdt'\lambda(t)\lambda(t')\langle\sigma^z_1(t)\sigma^z_1(t')\rangle_0,\nonumber\\
    &&+g^2_0\int_0^\tau dtdt'\lambda(t)\lambda(t')\langle\sigma^z_1(t)\sigma^z_1(t'+s)\rangle_0,\nonumber\\
    &&+\mc O(g^3_0).
    \label{char_func_Ape}
\end{eqnarray}
The first line in the above equation does not depend on $s$ because $E_0=0$. Thus, this term produces a $\delta(w)$ in $P(w,\tau)$. In addition, we see that $G(s,\tau)$ is completely determined by the autocorrelation function $\langle\sigma^z_1(t)\sigma^z_1(0)\rangle_0$. 

The second line of Eq. (\ref{char_func_Ape}) depends explicitly on $s$ through the autocorrelation function $\langle\sigma^z_1(t)\sigma^z_1(t'+s)\rangle_0$. Using the result derived in the main text [see discussion around Eq. (\ref{charac_func2})], we obtain for this term
\begin{eqnarray}
    \langle \sigma^z_1\rangle^2_0A(\tau)+4\sum_{\kappa,\kappa'}f_{\kappa\kappa'}\Lambda_{\kappa\kappa'}(\tau)e^{i(\Gamma_{\kappa}+\Gamma_{\kappa'})s},
    \label{char_func_Ape2}
\end{eqnarray}
where we defined $A(\tau)=g^2_0\left(\int_0^\tau dt\lambda(t)\right)^2$ and $\Lambda_{\kappa\kappa'}(\tau)=g^2_0|\int_0^\tau dt\lambda(t) e^{-i(\Gamma_{\kappa}+\Gamma_{\kappa'})t}|^2$. The first term in (\ref{char_func_Ape2}) also does not depend on $s$; only the second one does. This implies that the nontrivial part of $P(w,\tau)$ comes from the second term in Eq. (\ref{char_func_Ape2}).

The effects of the finite-time quench are fully encoded in the function $\Lambda_{\kappa\kappa'}(\tau)$. In general, this function modifies both the shape of the work distribution and the exponents associated with the ESs. The latter can be inferred directly from the behavior of $\Lambda_{\kappa\kappa'}(\tau)$. In particular, the low-energy ESs are controlled by the structure of $\Lambda_{\kappa\kappa'}(\tau)$ around $k=k'=0$ (in the presence of an edge mode with energy below $\Delta_-$, we have a one variable function). If the protocol is such that $\Lambda_{kk'}(\tau)|_{0,0}\sim1$, the ESs exponents remain those reported in Table~\ref{t1}. By contrast, if the protocol yields $\Lambda_{kk'}(\tau)|_{0,0}\sim k^{2m}k'^{2n}$, with $m,n\in\mathbb{N}$, one obtains $G(s\gg J^{-1},\tau)\sim e^{2i\Delta_{-}s}/s^{5+\nu}$, where $\nu=m+n$. As a consequence, the corresponding low-energy ES acquires the modified exponent $4+\nu$. The case $\nu=0$ reduces to the result in Eq. (\ref{edge_sing_deloc_deloc}).

Consider the case of a step-like quench, $\lambda(t)=1$ for $t\in (0,\tau)$ and zero otherwise. In this case we have
\begin{equation}
    \Lambda_{kk'}(\tau)=2g^2_0\frac{1-\cos\left[(\Gamma_{k}+\Gamma_{k'})\tau\right]}{(\Gamma_{k}+\Gamma_{k'})^2}.
\end{equation}
We can see that $\Lambda_{kk'}(\tau)\sim 1$ both around $k=k'=0$ and around $k=k'=\pi$ (including along the critical line $h=1$). Therefore, the ESs are governed by the same exponents listed in Table~\ref{t1}.

\bibliography{references.bib}

\begin{thebibliography}{62}%
\makeatletter
\providecommand \@ifxundefined [1]{%
 \@ifx{#1\undefined}
}%
\providecommand \@ifnum [1]{%
 \ifnum #1\expandafter \@firstoftwo
 \else \expandafter \@secondoftwo
 \fi
}%
\providecommand \@ifx [1]{%
 \ifx #1\expandafter \@firstoftwo
 \else \expandafter \@secondoftwo
 \fi
}%
\providecommand \natexlab [1]{#1}%
\providecommand \enquote  [1]{``#1''}%
\providecommand \bibnamefont  [1]{#1}%
\providecommand \bibfnamefont [1]{#1}%
\providecommand \citenamefont [1]{#1}%
\providecommand \href@noop [0]{\@secondoftwo}%
\providecommand \href [0]{\begingroup \@sanitize@url \@href}%
\providecommand \@href[1]{\@@startlink{#1}\@@href}%
\providecommand \@@href[1]{\endgroup#1\@@endlink}%
\providecommand \@sanitize@url [0]{\catcode `\\12\catcode `\$12\catcode `\&12\catcode `\#12\catcode `\^12\catcode `\_12\catcode `\%12\relax}%
\providecommand \@@startlink[1]{}%
\providecommand \@@endlink[0]{}%
\providecommand \url  [0]{\begingroup\@sanitize@url \@url }%
\providecommand \@url [1]{\endgroup\@href {#1}{\urlprefix }}%
\providecommand \urlprefix  [0]{URL }%
\providecommand \Eprint [0]{\href }%
\providecommand \doibase [0]{http://dx.doi.org/}%
\providecommand \selectlanguage [0]{\@gobble}%
\providecommand \bibinfo  [0]{\@secondoftwo}%
\providecommand \bibfield  [0]{\@secondoftwo}%
\providecommand \translation [1]{[#1]}%
\providecommand \BibitemOpen [0]{}%
\providecommand \bibitemStop [0]{}%
\providecommand \bibitemNoStop [0]{.\EOS\space}%
\providecommand \EOS [0]{\spacefactor3000\relax}%
\providecommand \BibitemShut  [1]{\csname bibitem#1\endcsname}%
\let\auto@bib@innerbib\@empty
\bibitem [{\citenamefont {Polkovnikov}\ \emph {et~al.}(2011)\citenamefont {Polkovnikov}, \citenamefont {Sengupta}, \citenamefont {Silva},\ and\ \citenamefont {Vengalattore}}]{RMP_2011}%
  \BibitemOpen
  \bibfield  {author} {\bibinfo {author} {\bibfnamefont {A.}~\bibnamefont {Polkovnikov}}, \bibinfo {author} {\bibfnamefont {K.}~\bibnamefont {Sengupta}}, \bibinfo {author} {\bibfnamefont {A.}~\bibnamefont {Silva}}, \ and\ \bibinfo {author} {\bibfnamefont {M.}~\bibnamefont {Vengalattore}},\ }\href {\doibase 10.1103/RevModPhys.83.863} {\bibfield  {journal} {\bibinfo  {journal} {Rev. Mod. Phys.}\ }\textbf {\bibinfo {volume} {83}},\ \bibinfo {pages} {863} (\bibinfo {year} {2011})}\BibitemShut {NoStop}%
\bibitem [{\citenamefont {Eisert}\ \emph {et~al.}(2015)\citenamefont {Eisert}, \citenamefont {Friesdorf},\ and\ \citenamefont {Gogolin}}]{Eisert2015}%
  \BibitemOpen
  \bibfield  {author} {\bibinfo {author} {\bibfnamefont {J.}~\bibnamefont {Eisert}}, \bibinfo {author} {\bibfnamefont {M.}~\bibnamefont {Friesdorf}}, \ and\ \bibinfo {author} {\bibfnamefont {C.}~\bibnamefont {Gogolin}},\ }\href {\doibase 10.1038/nphys3215} {\bibfield  {journal} {\bibinfo  {journal} {Nature Physics}\ }\textbf {\bibinfo {volume} {11}},\ \bibinfo {pages} {124} (\bibinfo {year} {2015})}\BibitemShut {NoStop}%
\bibitem [{\citenamefont {Gogolin}\ and\ \citenamefont {Eisert}(2016)}]{Gogolin_2016}%
  \BibitemOpen
  \bibfield  {author} {\bibinfo {author} {\bibfnamefont {C.}~\bibnamefont {Gogolin}}\ and\ \bibinfo {author} {\bibfnamefont {J.}~\bibnamefont {Eisert}},\ }\href {\doibase 10.1088/0034-4885/79/5/056001} {\bibfield  {journal} {\bibinfo  {journal} {Reports on Progress in Physics}\ }\textbf {\bibinfo {volume} {79}},\ \bibinfo {pages} {056001} (\bibinfo {year} {2016})}\BibitemShut {NoStop}%
\bibitem [{\citenamefont {Mori}\ \emph {et~al.}(2018)\citenamefont {Mori}, \citenamefont {Ikeda}, \citenamefont {Kaminishi},\ and\ \citenamefont {Ueda}}]{Mori_2018}%
  \BibitemOpen
  \bibfield  {author} {\bibinfo {author} {\bibfnamefont {T.}~\bibnamefont {Mori}}, \bibinfo {author} {\bibfnamefont {T.~N.}\ \bibnamefont {Ikeda}}, \bibinfo {author} {\bibfnamefont {E.}~\bibnamefont {Kaminishi}}, \ and\ \bibinfo {author} {\bibfnamefont {M.}~\bibnamefont {Ueda}},\ }\href {\doibase 10.1088/1361-6455/aabcdf} {\bibfield  {journal} {\bibinfo  {journal} {Journal of Physics B: Atomic, Molecular and Optical Physics}\ }\textbf {\bibinfo {volume} {51}},\ \bibinfo {pages} {112001} (\bibinfo {year} {2018})}\BibitemShut {NoStop}%
\bibitem [{\citenamefont {Abanin}\ \emph {et~al.}(2019)\citenamefont {Abanin}, \citenamefont {Altman}, \citenamefont {Bloch},\ and\ \citenamefont {Serbyn}}]{RMP_2019}%
  \BibitemOpen
  \bibfield  {author} {\bibinfo {author} {\bibfnamefont {D.~A.}\ \bibnamefont {Abanin}}, \bibinfo {author} {\bibfnamefont {E.}~\bibnamefont {Altman}}, \bibinfo {author} {\bibfnamefont {I.}~\bibnamefont {Bloch}}, \ and\ \bibinfo {author} {\bibfnamefont {M.}~\bibnamefont {Serbyn}},\ }\href {\doibase 10.1103/RevModPhys.91.021001} {\bibfield  {journal} {\bibinfo  {journal} {Rev. Mod. Phys.}\ }\textbf {\bibinfo {volume} {91}},\ \bibinfo {pages} {021001} (\bibinfo {year} {2019})}\BibitemShut {NoStop}%
\bibitem [{\citenamefont {Bloch}\ \emph {et~al.}(2008)\citenamefont {Bloch}, \citenamefont {Dalibard},\ and\ \citenamefont {Zwerger}}]{RMP_Bloch}%
  \BibitemOpen
  \bibfield  {author} {\bibinfo {author} {\bibfnamefont {I.}~\bibnamefont {Bloch}}, \bibinfo {author} {\bibfnamefont {J.}~\bibnamefont {Dalibard}}, \ and\ \bibinfo {author} {\bibfnamefont {W.}~\bibnamefont {Zwerger}},\ }\href {\doibase 10.1103/RevModPhys.80.885} {\bibfield  {journal} {\bibinfo  {journal} {Rev. Mod. Phys.}\ }\textbf {\bibinfo {volume} {80}},\ \bibinfo {pages} {885} (\bibinfo {year} {2008})}\BibitemShut {NoStop}%
\bibitem [{\citenamefont {Cazalilla}\ \emph {et~al.}(2011)\citenamefont {Cazalilla}, \citenamefont {Citro}, \citenamefont {Giamarchi}, \citenamefont {Orignac},\ and\ \citenamefont {Rigol}}]{RMP_Cazalilla_2011}%
  \BibitemOpen
  \bibfield  {author} {\bibinfo {author} {\bibfnamefont {M.~A.}\ \bibnamefont {Cazalilla}}, \bibinfo {author} {\bibfnamefont {R.}~\bibnamefont {Citro}}, \bibinfo {author} {\bibfnamefont {T.}~\bibnamefont {Giamarchi}}, \bibinfo {author} {\bibfnamefont {E.}~\bibnamefont {Orignac}}, \ and\ \bibinfo {author} {\bibfnamefont {M.}~\bibnamefont {Rigol}},\ }\href {\doibase 10.1103/RevModPhys.83.1405} {\bibfield  {journal} {\bibinfo  {journal} {Rev. Mod. Phys.}\ }\textbf {\bibinfo {volume} {83}},\ \bibinfo {pages} {1405} (\bibinfo {year} {2011})}\BibitemShut {NoStop}%
\bibitem [{\citenamefont {Trotzky}\ \emph {et~al.}(2012)\citenamefont {Trotzky}, \citenamefont {Chen}, \citenamefont {Flesch}, \citenamefont {McCulloch}, \citenamefont {Schollw{\"o}ck}, \citenamefont {Eisert},\ and\ \citenamefont {Bloch}}]{Trotzky2012}%
  \BibitemOpen
  \bibfield  {author} {\bibinfo {author} {\bibfnamefont {S.}~\bibnamefont {Trotzky}}, \bibinfo {author} {\bibfnamefont {Y.-A.}\ \bibnamefont {Chen}}, \bibinfo {author} {\bibfnamefont {A.}~\bibnamefont {Flesch}}, \bibinfo {author} {\bibfnamefont {I.~P.}\ \bibnamefont {McCulloch}}, \bibinfo {author} {\bibfnamefont {U.}~\bibnamefont {Schollw{\"o}ck}}, \bibinfo {author} {\bibfnamefont {J.}~\bibnamefont {Eisert}}, \ and\ \bibinfo {author} {\bibfnamefont {I.}~\bibnamefont {Bloch}},\ }\href {\doibase 10.1038/nphys2232} {\bibfield  {journal} {\bibinfo  {journal} {Nature Physics}\ }\textbf {\bibinfo {volume} {8}},\ \bibinfo {pages} {325} (\bibinfo {year} {2012})}\BibitemShut {NoStop}%
\bibitem [{\citenamefont {Aimet}\ \emph {et~al.}(2025)\citenamefont {Aimet}, \citenamefont {Tajik}, \citenamefont {Tournaire}, \citenamefont {Sch{\"u}ttelkopf}, \citenamefont {Sabino}, \citenamefont {Sotiriadis}, \citenamefont {Guarnieri}, \citenamefont {Schmiedmayer},\ and\ \citenamefont {Eisert}}]{Aimet2025}%
  \BibitemOpen
  \bibfield  {author} {\bibinfo {author} {\bibfnamefont {S.}~\bibnamefont {Aimet}}, \bibinfo {author} {\bibfnamefont {M.}~\bibnamefont {Tajik}}, \bibinfo {author} {\bibfnamefont {G.}~\bibnamefont {Tournaire}}, \bibinfo {author} {\bibfnamefont {P.}~\bibnamefont {Sch{\"u}ttelkopf}}, \bibinfo {author} {\bibfnamefont {J.}~\bibnamefont {Sabino}}, \bibinfo {author} {\bibfnamefont {S.}~\bibnamefont {Sotiriadis}}, \bibinfo {author} {\bibfnamefont {G.}~\bibnamefont {Guarnieri}}, \bibinfo {author} {\bibfnamefont {J.}~\bibnamefont {Schmiedmayer}}, \ and\ \bibinfo {author} {\bibfnamefont {J.}~\bibnamefont {Eisert}},\ }\href {\doibase 10.1038/s41567-025-02930-9} {\bibfield  {journal} {\bibinfo  {journal} {Nature Physics}\ }\textbf {\bibinfo {volume} {21}},\ \bibinfo {pages} {1326} (\bibinfo {year} {2025})}\BibitemShut {NoStop}%
\bibitem [{\citenamefont {Cetina}\ \emph {et~al.}(2016)\citenamefont {Cetina}, \citenamefont {Jag}, \citenamefont {Lous}, \citenamefont {Fritsche}, \citenamefont {Walraven}, \citenamefont {Grimm}, \citenamefont {Levinsen}, \citenamefont {Parish}, \citenamefont {Schmidt}, \citenamefont {Knap},\ and\ \citenamefont {Demler}}]{doi:10.1126/science.aaf5134}%
  \BibitemOpen
  \bibfield  {author} {\bibinfo {author} {\bibfnamefont {M.}~\bibnamefont {Cetina}}, \bibinfo {author} {\bibfnamefont {M.}~\bibnamefont {Jag}}, \bibinfo {author} {\bibfnamefont {R.~S.}\ \bibnamefont {Lous}}, \bibinfo {author} {\bibfnamefont {I.}~\bibnamefont {Fritsche}}, \bibinfo {author} {\bibfnamefont {J.~T.~M.}\ \bibnamefont {Walraven}}, \bibinfo {author} {\bibfnamefont {R.}~\bibnamefont {Grimm}}, \bibinfo {author} {\bibfnamefont {J.}~\bibnamefont {Levinsen}}, \bibinfo {author} {\bibfnamefont {M.~M.}\ \bibnamefont {Parish}}, \bibinfo {author} {\bibfnamefont {R.}~\bibnamefont {Schmidt}}, \bibinfo {author} {\bibfnamefont {M.}~\bibnamefont {Knap}}, \ and\ \bibinfo {author} {\bibfnamefont {E.}~\bibnamefont {Demler}},\ }\href {\doibase 10.1126/science.aaf5134} {\bibfield  {journal} {\bibinfo  {journal} {Science}\ }\textbf {\bibinfo {volume} {354}},\ \bibinfo {pages} {96} (\bibinfo {year} {2016})}\BibitemShut {NoStop}%
\bibitem [{\citenamefont {Evered}\ \emph {et~al.}(2025)\citenamefont {Evered}, \citenamefont {Kalinowski}, \citenamefont {Geim}, \citenamefont {Manovitz}, \citenamefont {Bluvstein}, \citenamefont {Li}, \citenamefont {Maskara}, \citenamefont {Zhou}, \citenamefont {Ebadi}, \citenamefont {Xu}, \citenamefont {Campo}, \citenamefont {Cain}, \citenamefont {Ostermann}, \citenamefont {Yelin}, \citenamefont {Sachdev}, \citenamefont {Greiner}, \citenamefont {Vuleti{\'{c}}},\ and\ \citenamefont {Lukin}}]{Evered2025}%
  \BibitemOpen
  \bibfield  {author} {\bibinfo {author} {\bibfnamefont {S.~J.}\ \bibnamefont {Evered}}, \bibinfo {author} {\bibfnamefont {M.}~\bibnamefont {Kalinowski}}, \bibinfo {author} {\bibfnamefont {A.~A.}\ \bibnamefont {Geim}}, \bibinfo {author} {\bibfnamefont {T.}~\bibnamefont {Manovitz}}, \bibinfo {author} {\bibfnamefont {D.}~\bibnamefont {Bluvstein}}, \bibinfo {author} {\bibfnamefont {S.~H.}\ \bibnamefont {Li}}, \bibinfo {author} {\bibfnamefont {N.}~\bibnamefont {Maskara}}, \bibinfo {author} {\bibfnamefont {H.}~\bibnamefont {Zhou}}, \bibinfo {author} {\bibfnamefont {S.}~\bibnamefont {Ebadi}}, \bibinfo {author} {\bibfnamefont {M.}~\bibnamefont {Xu}}, \bibinfo {author} {\bibfnamefont {J.}~\bibnamefont {Campo}}, \bibinfo {author} {\bibfnamefont {M.}~\bibnamefont {Cain}}, \bibinfo {author} {\bibfnamefont {S.}~\bibnamefont {Ostermann}}, \bibinfo {author} {\bibfnamefont {S.~F.}\ \bibnamefont {Yelin}}, \bibinfo {author} {\bibfnamefont {S.}~\bibnamefont {Sachdev}}, \bibinfo {author} {\bibfnamefont {M.}~\bibnamefont
  {Greiner}}, \bibinfo {author} {\bibfnamefont {V.}~\bibnamefont {Vuleti{\'{c}}}}, \ and\ \bibinfo {author} {\bibfnamefont {M.~D.}\ \bibnamefont {Lukin}},\ }\href {\doibase 10.1038/s41586-025-09475-0} {\bibfield  {journal} {\bibinfo  {journal} {Nature}\ }\textbf {\bibinfo {volume} {645}},\ \bibinfo {pages} {341} (\bibinfo {year} {2025})}\BibitemShut {NoStop}%
\bibitem [{\citenamefont {Bluvstein}\ \emph {et~al.}(2021)\citenamefont {Bluvstein}, \citenamefont {Omran}, \citenamefont {Levine}, \citenamefont {Keesling}, \citenamefont {Semeghini}, \citenamefont {Ebadi}, \citenamefont {Wang}, \citenamefont {Michailidis}, \citenamefont {Maskara}, \citenamefont {Ho}, \citenamefont {Choi}, \citenamefont {Serbyn}, \citenamefont {Greiner}, \citenamefont {Vuleti\'{c}},\ and\ \citenamefont {Lukin}}]{Rydn_exp_2021}%
  \BibitemOpen
  \bibfield  {author} {\bibinfo {author} {\bibfnamefont {D.}~\bibnamefont {Bluvstein}}, \bibinfo {author} {\bibfnamefont {A.}~\bibnamefont {Omran}}, \bibinfo {author} {\bibfnamefont {H.}~\bibnamefont {Levine}}, \bibinfo {author} {\bibfnamefont {A.}~\bibnamefont {Keesling}}, \bibinfo {author} {\bibfnamefont {G.}~\bibnamefont {Semeghini}}, \bibinfo {author} {\bibfnamefont {S.}~\bibnamefont {Ebadi}}, \bibinfo {author} {\bibfnamefont {T.~T.}\ \bibnamefont {Wang}}, \bibinfo {author} {\bibfnamefont {A.~A.}\ \bibnamefont {Michailidis}}, \bibinfo {author} {\bibfnamefont {N.}~\bibnamefont {Maskara}}, \bibinfo {author} {\bibfnamefont {W.~W.}\ \bibnamefont {Ho}}, \bibinfo {author} {\bibfnamefont {S.}~\bibnamefont {Choi}}, \bibinfo {author} {\bibfnamefont {M.}~\bibnamefont {Serbyn}}, \bibinfo {author} {\bibfnamefont {M.}~\bibnamefont {Greiner}}, \bibinfo {author} {\bibfnamefont {V.}~\bibnamefont {Vuleti\'{c}}}, \ and\ \bibinfo {author} {\bibfnamefont {M.~D.}\ \bibnamefont {Lukin}},\ }\href {\doibase
  10.1126/science.abg2530} {\bibfield  {journal} {\bibinfo  {journal} {Science}\ }\textbf {\bibinfo {volume} {371}},\ \bibinfo {pages} {1355} (\bibinfo {year} {2021})}\BibitemShut {NoStop}%
\bibitem [{\citenamefont {Dumitrescu}\ \emph {et~al.}(2022)\citenamefont {Dumitrescu}, \citenamefont {Bohnet}, \citenamefont {Gaebler}, \citenamefont {Hankin}, \citenamefont {Hayes}, \citenamefont {Kumar}, \citenamefont {Neyenhuis}, \citenamefont {Vasseur},\ and\ \citenamefont {Potter}}]{Dumitrescu2022}%
  \BibitemOpen
  \bibfield  {author} {\bibinfo {author} {\bibfnamefont {P.~T.}\ \bibnamefont {Dumitrescu}}, \bibinfo {author} {\bibfnamefont {J.~G.}\ \bibnamefont {Bohnet}}, \bibinfo {author} {\bibfnamefont {J.~P.}\ \bibnamefont {Gaebler}}, \bibinfo {author} {\bibfnamefont {A.}~\bibnamefont {Hankin}}, \bibinfo {author} {\bibfnamefont {D.}~\bibnamefont {Hayes}}, \bibinfo {author} {\bibfnamefont {A.}~\bibnamefont {Kumar}}, \bibinfo {author} {\bibfnamefont {B.}~\bibnamefont {Neyenhuis}}, \bibinfo {author} {\bibfnamefont {R.}~\bibnamefont {Vasseur}}, \ and\ \bibinfo {author} {\bibfnamefont {A.~C.}\ \bibnamefont {Potter}},\ }\href {\doibase 10.1038/s41586-022-04853-4} {\bibfield  {journal} {\bibinfo  {journal} {Nature}\ }\textbf {\bibinfo {volume} {607}},\ \bibinfo {pages} {463} (\bibinfo {year} {2022})}\BibitemShut {NoStop}%
\bibitem [{\citenamefont {Mitra}(2018)}]{Mitra_Quench_2018}%
  \BibitemOpen
  \bibfield  {author} {\bibinfo {author} {\bibfnamefont {A.}~\bibnamefont {Mitra}},\ }\href {\doibase https://doi.org/10.1146/annurev-conmatphys-031016-025451} {\bibfield  {journal} {\bibinfo  {journal} {Annual Review of Condensed Matter Physics}\ }\textbf {\bibinfo {volume} {9}},\ \bibinfo {pages} {245} (\bibinfo {year} {2018})}\BibitemShut {NoStop}%
\bibitem [{\citenamefont {Kinoshita}\ \emph {et~al.}(2006)\citenamefont {Kinoshita}, \citenamefont {Wenger},\ and\ \citenamefont {Weiss}}]{Kinoshita2006}%
  \BibitemOpen
  \bibfield  {author} {\bibinfo {author} {\bibfnamefont {T.}~\bibnamefont {Kinoshita}}, \bibinfo {author} {\bibfnamefont {T.}~\bibnamefont {Wenger}}, \ and\ \bibinfo {author} {\bibfnamefont {D.~S.}\ \bibnamefont {Weiss}},\ }\href {\doibase 10.1038/nature04693} {\bibfield  {journal} {\bibinfo  {journal} {Nature}\ }\textbf {\bibinfo {volume} {440}},\ \bibinfo {pages} {900} (\bibinfo {year} {2006})}\BibitemShut {NoStop}%
\bibitem [{\citenamefont {Gring}\ \emph {et~al.}(2012)\citenamefont {Gring}, \citenamefont {Kuhnert}, \citenamefont {Langen}, \citenamefont {Kitagawa}, \citenamefont {Rauer}, \citenamefont {Schreitl}, \citenamefont {Mazets}, \citenamefont {Smith}, \citenamefont {Demler},\ and\ \citenamefont {Schmiedmayer}}]{Pre_Science_2012}%
  \BibitemOpen
  \bibfield  {author} {\bibinfo {author} {\bibfnamefont {M.}~\bibnamefont {Gring}}, \bibinfo {author} {\bibfnamefont {M.}~\bibnamefont {Kuhnert}}, \bibinfo {author} {\bibfnamefont {T.}~\bibnamefont {Langen}}, \bibinfo {author} {\bibfnamefont {T.}~\bibnamefont {Kitagawa}}, \bibinfo {author} {\bibfnamefont {B.}~\bibnamefont {Rauer}}, \bibinfo {author} {\bibfnamefont {M.}~\bibnamefont {Schreitl}}, \bibinfo {author} {\bibfnamefont {I.}~\bibnamefont {Mazets}}, \bibinfo {author} {\bibfnamefont {D.~A.}\ \bibnamefont {Smith}}, \bibinfo {author} {\bibfnamefont {E.}~\bibnamefont {Demler}}, \ and\ \bibinfo {author} {\bibfnamefont {J.}~\bibnamefont {Schmiedmayer}},\ }\href {\doibase 10.1126/science.1224953} {\bibfield  {journal} {\bibinfo  {journal} {Science}\ }\textbf {\bibinfo {volume} {337}},\ \bibinfo {pages} {1318} (\bibinfo {year} {2012})}\BibitemShut {NoStop}%
\bibitem [{\citenamefont {Kaufman}\ \emph {et~al.}(2016)\citenamefont {Kaufman}, \citenamefont {Tai}, \citenamefont {Lukin}, \citenamefont {Rispoli}, \citenamefont {Schittko}, \citenamefont {Preiss},\ and\ \citenamefont {Greiner}}]{Pre_Science_2016}%
  \BibitemOpen
  \bibfield  {author} {\bibinfo {author} {\bibfnamefont {A.~M.}\ \bibnamefont {Kaufman}}, \bibinfo {author} {\bibfnamefont {M.~E.}\ \bibnamefont {Tai}}, \bibinfo {author} {\bibfnamefont {A.}~\bibnamefont {Lukin}}, \bibinfo {author} {\bibfnamefont {M.}~\bibnamefont {Rispoli}}, \bibinfo {author} {\bibfnamefont {R.}~\bibnamefont {Schittko}}, \bibinfo {author} {\bibfnamefont {P.~M.}\ \bibnamefont {Preiss}}, \ and\ \bibinfo {author} {\bibfnamefont {M.}~\bibnamefont {Greiner}},\ }\href {\doibase 10.1126/science.aaf6725} {\bibfield  {journal} {\bibinfo  {journal} {Science}\ }\textbf {\bibinfo {volume} {353}},\ \bibinfo {pages} {794} (\bibinfo {year} {2016})},\ \Eprint {http://arxiv.org/abs/https://www.science.org/doi/pdf/10.1126/science.aaf6725} {https://www.science.org/doi/pdf/10.1126/science.aaf6725} \BibitemShut {NoStop}%
\bibitem [{\citenamefont {Cheneau}\ \emph {et~al.}(2012)\citenamefont {Cheneau}, \citenamefont {Barmettler}, \citenamefont {Poletti}, \citenamefont {Endres}, \citenamefont {Schau{\ss}}, \citenamefont {Fukuhara}, \citenamefont {Gross}, \citenamefont {Bloch}, \citenamefont {Kollath},\ and\ \citenamefont {Kuhr}}]{Cheneau_2012}%
  \BibitemOpen
  \bibfield  {author} {\bibinfo {author} {\bibfnamefont {M.}~\bibnamefont {Cheneau}}, \bibinfo {author} {\bibfnamefont {P.}~\bibnamefont {Barmettler}}, \bibinfo {author} {\bibfnamefont {D.}~\bibnamefont {Poletti}}, \bibinfo {author} {\bibfnamefont {M.}~\bibnamefont {Endres}}, \bibinfo {author} {\bibfnamefont {P.}~\bibnamefont {Schau{\ss}}}, \bibinfo {author} {\bibfnamefont {T.}~\bibnamefont {Fukuhara}}, \bibinfo {author} {\bibfnamefont {C.}~\bibnamefont {Gross}}, \bibinfo {author} {\bibfnamefont {I.}~\bibnamefont {Bloch}}, \bibinfo {author} {\bibfnamefont {C.}~\bibnamefont {Kollath}}, \ and\ \bibinfo {author} {\bibfnamefont {S.}~\bibnamefont {Kuhr}},\ }\href {\doibase 10.1038/nature10748} {\bibfield  {journal} {\bibinfo  {journal} {Nature}\ }\textbf {\bibinfo {volume} {481}},\ \bibinfo {pages} {484} (\bibinfo {year} {2012})}\BibitemShut {NoStop}%
\bibitem [{\citenamefont {Langen}\ \emph {et~al.}(2013)\citenamefont {Langen}, \citenamefont {Geiger}, \citenamefont {Kuhnert}, \citenamefont {Rauer},\ and\ \citenamefont {Schmiedmayer}}]{Langen2013}%
  \BibitemOpen
  \bibfield  {author} {\bibinfo {author} {\bibfnamefont {T.}~\bibnamefont {Langen}}, \bibinfo {author} {\bibfnamefont {R.}~\bibnamefont {Geiger}}, \bibinfo {author} {\bibfnamefont {M.}~\bibnamefont {Kuhnert}}, \bibinfo {author} {\bibfnamefont {B.}~\bibnamefont {Rauer}}, \ and\ \bibinfo {author} {\bibfnamefont {J.}~\bibnamefont {Schmiedmayer}},\ }\href {\doibase 10.1038/nphys2739} {\bibfield  {journal} {\bibinfo  {journal} {Nature Physics}\ }\textbf {\bibinfo {volume} {9}},\ \bibinfo {pages} {640} (\bibinfo {year} {2013})}\BibitemShut {NoStop}%
\bibitem [{\citenamefont {Liang}\ \emph {et~al.}(2025)\citenamefont {Liang}, \citenamefont {Yue}, \citenamefont {Chao}, \citenamefont {Hua}, \citenamefont {Lin}, \citenamefont {Tey},\ and\ \citenamefont {You}}]{w1cp-l5vq}%
  \BibitemOpen
  \bibfield  {author} {\bibinfo {author} {\bibfnamefont {X.}~\bibnamefont {Liang}}, \bibinfo {author} {\bibfnamefont {Z.}~\bibnamefont {Yue}}, \bibinfo {author} {\bibfnamefont {Y.-X.}\ \bibnamefont {Chao}}, \bibinfo {author} {\bibfnamefont {Z.-X.}\ \bibnamefont {Hua}}, \bibinfo {author} {\bibfnamefont {Y.}~\bibnamefont {Lin}}, \bibinfo {author} {\bibfnamefont {M.~K.}\ \bibnamefont {Tey}}, \ and\ \bibinfo {author} {\bibfnamefont {L.}~\bibnamefont {You}},\ }\href {\doibase 10.1103/w1cp-l5vq} {\bibfield  {journal} {\bibinfo  {journal} {Phys. Rev. Lett.}\ }\textbf {\bibinfo {volume} {135}},\ \bibinfo {pages} {050201} (\bibinfo {year} {2025})}\BibitemShut {NoStop}%
\bibitem [{\citenamefont {Yarkoni}\ \emph {et~al.}(2022)\citenamefont {Yarkoni}, \citenamefont {Raponi}, \citenamefont {Bäck},\ and\ \citenamefont {Schmitt}}]{Yarkoni_2022}%
  \BibitemOpen
  \bibfield  {author} {\bibinfo {author} {\bibfnamefont {S.}~\bibnamefont {Yarkoni}}, \bibinfo {author} {\bibfnamefont {E.}~\bibnamefont {Raponi}}, \bibinfo {author} {\bibfnamefont {T.}~\bibnamefont {Bäck}}, \ and\ \bibinfo {author} {\bibfnamefont {S.}~\bibnamefont {Schmitt}},\ }\href {\doibase 10.1088/1361-6633/ac8c54} {\bibfield  {journal} {\bibinfo  {journal} {Reports on Progress in Physics}\ }\textbf {\bibinfo {volume} {85}},\ \bibinfo {pages} {104001} (\bibinfo {year} {2022})}\BibitemShut {NoStop}%
\bibitem [{\citenamefont {Silva}(2008)}]{PRL_Ale_2008}%
  \BibitemOpen
  \bibfield  {author} {\bibinfo {author} {\bibfnamefont {A.}~\bibnamefont {Silva}},\ }\href {\doibase 10.1103/PhysRevLett.101.120603} {\bibfield  {journal} {\bibinfo  {journal} {Phys. Rev. Lett.}\ }\textbf {\bibinfo {volume} {101}},\ \bibinfo {pages} {120603} (\bibinfo {year} {2008})}\BibitemShut {NoStop}%
\bibitem [{\citenamefont {Sindona}\ \emph {et~al.}(2014)\citenamefont {Sindona}, \citenamefont {Goold}, \citenamefont {Lo~Gullo},\ and\ \citenamefont {Plastina}}]{Sindona_2014}%
  \BibitemOpen
  \bibfield  {author} {\bibinfo {author} {\bibfnamefont {A.}~\bibnamefont {Sindona}}, \bibinfo {author} {\bibfnamefont {J.}~\bibnamefont {Goold}}, \bibinfo {author} {\bibfnamefont {N.}~\bibnamefont {Lo~Gullo}}, \ and\ \bibinfo {author} {\bibfnamefont {F.}~\bibnamefont {Plastina}},\ }\href {\doibase 10.1088/1367-2630/16/4/045013} {\bibfield  {journal} {\bibinfo  {journal} {New Journal of Physics}\ }\textbf {\bibinfo {volume} {16}},\ \bibinfo {pages} {045013} (\bibinfo {year} {2014})}\BibitemShut {NoStop}%
\bibitem [{\citenamefont {Landi}\ and\ \citenamefont {Karevski}(2016)}]{PRE_Landi_2016}%
  \BibitemOpen
  \bibfield  {author} {\bibinfo {author} {\bibfnamefont {G.~T.}\ \bibnamefont {Landi}}\ and\ \bibinfo {author} {\bibfnamefont {D.}~\bibnamefont {Karevski}},\ }\href {\doibase 10.1103/PhysRevE.93.032122} {\bibfield  {journal} {\bibinfo  {journal} {Phys. Rev. E}\ }\textbf {\bibinfo {volume} {93}},\ \bibinfo {pages} {032122} (\bibinfo {year} {2016})}\BibitemShut {NoStop}%
\bibitem [{\citenamefont {Goold}\ \emph {et~al.}(2018)\citenamefont {Goold}, \citenamefont {Plastina}, \citenamefont {Gambassi},\ and\ \citenamefont {Silva}}]{Goold2018}%
  \BibitemOpen
  \bibfield  {author} {\bibinfo {author} {\bibfnamefont {J.}~\bibnamefont {Goold}}, \bibinfo {author} {\bibfnamefont {F.}~\bibnamefont {Plastina}}, \bibinfo {author} {\bibfnamefont {A.}~\bibnamefont {Gambassi}}, \ and\ \bibinfo {author} {\bibfnamefont {A.}~\bibnamefont {Silva}},\ }\enquote {\bibinfo {title} {The role of quantum work statistics in many-body physics},}\ in\ \href {\doibase 10.1007/978-3-319-99046-0_13} {\emph {\bibinfo {booktitle} {Thermodynamics in the Quantum Regime: Fundamental Aspects and New Directions}}},\ \bibinfo {editor} {edited by\ \bibinfo {editor} {\bibfnamefont {F.}~\bibnamefont {Binder}}, \bibinfo {editor} {\bibfnamefont {L.~A.}\ \bibnamefont {Correa}}, \bibinfo {editor} {\bibfnamefont {C.}~\bibnamefont {Gogolin}}, \bibinfo {editor} {\bibfnamefont {J.}~\bibnamefont {Anders}}, \ and\ \bibinfo {editor} {\bibfnamefont {G.}~\bibnamefont {Adesso}}}\ (\bibinfo  {publisher} {Springer International Publishing},\ \bibinfo {address} {Cham},\ \bibinfo {year} {2018})\ pp.\ \bibinfo {pages}
  {317--336}\BibitemShut {NoStop}%
\bibitem [{\citenamefont {Zawadzki}\ \emph {et~al.}(2020)\citenamefont {Zawadzki}, \citenamefont {Serra},\ and\ \citenamefont {D'Amico}}]{PRR_Krissia_2020}%
  \BibitemOpen
  \bibfield  {author} {\bibinfo {author} {\bibfnamefont {K.}~\bibnamefont {Zawadzki}}, \bibinfo {author} {\bibfnamefont {R.~M.}\ \bibnamefont {Serra}}, \ and\ \bibinfo {author} {\bibfnamefont {I.}~\bibnamefont {D'Amico}},\ }\href {\doibase 10.1103/PhysRevResearch.2.033167} {\bibfield  {journal} {\bibinfo  {journal} {Phys. Rev. Res.}\ }\textbf {\bibinfo {volume} {2}},\ \bibinfo {pages} {033167} (\bibinfo {year} {2020})}\BibitemShut {NoStop}%
\bibitem [{\citenamefont {Kiely}\ \emph {et~al.}(2023)\citenamefont {Kiely}, \citenamefont {O'Connor}, \citenamefont {Fogarty}, \citenamefont {Landi},\ and\ \citenamefont {Campbell}}]{PRR_Kiely_2023}%
  \BibitemOpen
  \bibfield  {author} {\bibinfo {author} {\bibfnamefont {A.}~\bibnamefont {Kiely}}, \bibinfo {author} {\bibfnamefont {E.}~\bibnamefont {O'Connor}}, \bibinfo {author} {\bibfnamefont {T.}~\bibnamefont {Fogarty}}, \bibinfo {author} {\bibfnamefont {G.~T.}\ \bibnamefont {Landi}}, \ and\ \bibinfo {author} {\bibfnamefont {S.}~\bibnamefont {Campbell}},\ }\href {\doibase 10.1103/PhysRevResearch.5.L022010} {\bibfield  {journal} {\bibinfo  {journal} {Phys. Rev. Res.}\ }\textbf {\bibinfo {volume} {5}},\ \bibinfo {pages} {L022010} (\bibinfo {year} {2023})}\BibitemShut {NoStop}%
\bibitem [{\citenamefont {Guarnieri}\ \emph {et~al.}(2024)\citenamefont {Guarnieri}, \citenamefont {Eisert},\ and\ \citenamefont {Miller}}]{Gen_LRT_PRL_2024}%
  \BibitemOpen
  \bibfield  {author} {\bibinfo {author} {\bibfnamefont {G.}~\bibnamefont {Guarnieri}}, \bibinfo {author} {\bibfnamefont {J.}~\bibnamefont {Eisert}}, \ and\ \bibinfo {author} {\bibfnamefont {H.~J.~D.}\ \bibnamefont {Miller}},\ }\href {\doibase 10.1103/PhysRevLett.133.070405} {\bibfield  {journal} {\bibinfo  {journal} {Phys. Rev. Lett.}\ }\textbf {\bibinfo {volume} {133}},\ \bibinfo {pages} {070405} (\bibinfo {year} {2024})}\BibitemShut {NoStop}%
\bibitem [{\citenamefont {D\'ora}\ and\ \citenamefont {Moca}(2024)}]{PRB_Dora_2024}%
  \BibitemOpen
  \bibfield  {author} {\bibinfo {author} {\bibfnamefont {B.}~\bibnamefont {D\'ora}}\ and\ \bibinfo {author} {\bibfnamefont {C.~P.}\ \bibnamefont {Moca}},\ }\href {\doibase 10.1103/PhysRevB.110.L121116} {\bibfield  {journal} {\bibinfo  {journal} {Phys. Rev. B}\ }\textbf {\bibinfo {volume} {110}},\ \bibinfo {pages} {L121116} (\bibinfo {year} {2024})}\BibitemShut {NoStop}%
\bibitem [{\citenamefont {Ma}\ \emph {et~al.}(2025)\citenamefont {Ma}, \citenamefont {Mitchell},\ and\ \citenamefont {Sela}}]{vn83-mt2v}%
  \BibitemOpen
  \bibfield  {author} {\bibinfo {author} {\bibfnamefont {Z.}~\bibnamefont {Ma}}, \bibinfo {author} {\bibfnamefont {A.~K.}\ \bibnamefont {Mitchell}}, \ and\ \bibinfo {author} {\bibfnamefont {E.}~\bibnamefont {Sela}},\ }\href {\doibase 10.1103/vn83-mt2v} {\bibfield  {journal} {\bibinfo  {journal} {Phys. Rev. Lett.}\ }\textbf {\bibinfo {volume} {135}},\ \bibinfo {pages} {130402} (\bibinfo {year} {2025})}\BibitemShut {NoStop}%
\bibitem [{\citenamefont {Murphy}\ \emph {et~al.}(2025)\citenamefont {Murphy}, \citenamefont {Kiely}, \citenamefont {D'Amico},\ and\ \citenamefont {Campbell}}]{96cm-ktcb}%
  \BibitemOpen
  \bibfield  {author} {\bibinfo {author} {\bibfnamefont {D.}~\bibnamefont {Murphy}}, \bibinfo {author} {\bibfnamefont {A.}~\bibnamefont {Kiely}}, \bibinfo {author} {\bibfnamefont {I.}~\bibnamefont {D'Amico}}, \ and\ \bibinfo {author} {\bibfnamefont {S.}~\bibnamefont {Campbell}},\ }\href {\doibase 10.1103/96cm-ktcb} {\bibfield  {journal} {\bibinfo  {journal} {Phys. Rev. A}\ }\textbf {\bibinfo {volume} {112}},\ \bibinfo {pages} {052214} (\bibinfo {year} {2025})}\BibitemShut {NoStop}%
\bibitem [{\citenamefont {Plastina}\ and\ \citenamefont {Apollaro}(2007)}]{PRL_Tony_2007}%
  \BibitemOpen
  \bibfield  {author} {\bibinfo {author} {\bibfnamefont {F.}~\bibnamefont {Plastina}}\ and\ \bibinfo {author} {\bibfnamefont {T.~J.~G.}\ \bibnamefont {Apollaro}},\ }\href {\doibase 10.1103/PhysRevLett.99.177210} {\bibfield  {journal} {\bibinfo  {journal} {Phys. Rev. Lett.}\ }\textbf {\bibinfo {volume} {99}},\ \bibinfo {pages} {177210} (\bibinfo {year} {2007})}\BibitemShut {NoStop}%
\bibitem [{\citenamefont {Francica}\ \emph {et~al.}(2016)\citenamefont {Francica}, \citenamefont {Apollaro}, \citenamefont {Lo~Gullo},\ and\ \citenamefont {Plastina}}]{PRB_main_ref_2016}%
  \BibitemOpen
  \bibfield  {author} {\bibinfo {author} {\bibfnamefont {G.}~\bibnamefont {Francica}}, \bibinfo {author} {\bibfnamefont {T.~J.~G.}\ \bibnamefont {Apollaro}}, \bibinfo {author} {\bibfnamefont {N.}~\bibnamefont {Lo~Gullo}}, \ and\ \bibinfo {author} {\bibfnamefont {F.}~\bibnamefont {Plastina}},\ }\href {\doibase 10.1103/PhysRevB.94.245103} {\bibfield  {journal} {\bibinfo  {journal} {Phys. Rev. B}\ }\textbf {\bibinfo {volume} {94}},\ \bibinfo {pages} {245103} (\bibinfo {year} {2016})}\BibitemShut {NoStop}%
\bibitem [{\citenamefont {M\"{u}ller}\ and\ \citenamefont {Nersesyan}(2016)}]{MULLER2016482}%
  \BibitemOpen
  \bibfield  {author} {\bibinfo {author} {\bibfnamefont {M.}~\bibnamefont {M\"{u}ller}}\ and\ \bibinfo {author} {\bibfnamefont {A.~A.}\ \bibnamefont {Nersesyan}},\ }\href {\doibase https://doi.org/10.1016/j.aop.2016.07.025} {\bibfield  {journal} {\bibinfo  {journal} {Annals of Physics}\ }\textbf {\bibinfo {volume} {372}},\ \bibinfo {pages} {482} (\bibinfo {year} {2016})}\BibitemShut {NoStop}%
\bibitem [{\citenamefont {Tang}\ \emph {et~al.}(2025)\citenamefont {Tang}, \citenamefont {Kattel}, \citenamefont {Pixley},\ and\ \citenamefont {Andrei}}]{t78z-pyfr}%
  \BibitemOpen
  \bibfield  {author} {\bibinfo {author} {\bibfnamefont {Y.}~\bibnamefont {Tang}}, \bibinfo {author} {\bibfnamefont {P.}~\bibnamefont {Kattel}}, \bibinfo {author} {\bibfnamefont {J.~H.}\ \bibnamefont {Pixley}}, \ and\ \bibinfo {author} {\bibfnamefont {N.}~\bibnamefont {Andrei}},\ }\href {\doibase 10.1103/t78z-pyfr} {\bibfield  {journal} {\bibinfo  {journal} {Phys. Rev. B}\ }\textbf {\bibinfo {volume} {112}},\ \bibinfo {pages} {L020303} (\bibinfo {year} {2025})}\BibitemShut {NoStop}%
\bibitem [{\citenamefont {Kitaev}(2001)}]{Kitaev_2001}%
  \BibitemOpen
  \bibfield  {author} {\bibinfo {author} {\bibfnamefont {A.~Y.}\ \bibnamefont {Kitaev}},\ }\href {\doibase 10.1070/1063-7869/44/10S/S29} {\bibfield  {journal} {\bibinfo  {journal} {Physics-Uspekhi}\ }\textbf {\bibinfo {volume} {44}},\ \bibinfo {pages} {131} (\bibinfo {year} {2001})}\BibitemShut {NoStop}%
\bibitem [{\citenamefont {Alicea}(2012)}]{Alicea_2012}%
  \BibitemOpen
  \bibfield  {author} {\bibinfo {author} {\bibfnamefont {J.}~\bibnamefont {Alicea}},\ }\href {\doibase 10.1088/0034-4885/75/7/076501} {\bibfield  {journal} {\bibinfo  {journal} {Reports on Progress in Physics}\ }\textbf {\bibinfo {volume} {75}},\ \bibinfo {pages} {076501} (\bibinfo {year} {2012})}\BibitemShut {NoStop}%
\bibitem [{\citenamefont {Moore}\ \emph {et~al.}(2018)\citenamefont {Moore}, \citenamefont {Stanescu},\ and\ \citenamefont {Tewari}}]{CMoore_PRB_2018}%
  \BibitemOpen
  \bibfield  {author} {\bibinfo {author} {\bibfnamefont {C.}~\bibnamefont {Moore}}, \bibinfo {author} {\bibfnamefont {T.~D.}\ \bibnamefont {Stanescu}}, \ and\ \bibinfo {author} {\bibfnamefont {S.}~\bibnamefont {Tewari}},\ }\href {\doibase 10.1103/PhysRevB.97.165302} {\bibfield  {journal} {\bibinfo  {journal} {Phys. Rev. B}\ }\textbf {\bibinfo {volume} {97}},\ \bibinfo {pages} {165302} (\bibinfo {year} {2018})}\BibitemShut {NoStop}%
\bibitem [{\citenamefont {Garc{\'i}a~Rodr{\'i}guez}\ and\ \citenamefont {Cabrera}(2018)}]{GarciaRodriguez2018}%
  \BibitemOpen
  \bibfield  {author} {\bibinfo {author} {\bibfnamefont {A.~O.}\ \bibnamefont {Garc{\'i}a~Rodr{\'i}guez}}\ and\ \bibinfo {author} {\bibfnamefont {G.~G.}\ \bibnamefont {Cabrera}},\ }\href {\doibase 10.1140/epjst/e2018-00095-7} {\bibfield  {journal} {\bibinfo  {journal} {The European Physical Journal Special Topics}\ }\textbf {\bibinfo {volume} {227}},\ \bibinfo {pages} {301} (\bibinfo {year} {2018})}\BibitemShut {NoStop}%
\bibitem [{\citenamefont {Javed}\ \emph {et~al.}(2023)\citenamefont {Javed}, \citenamefont {Marino}, \citenamefont {Oganesyan},\ and\ \citenamefont {Kolodrubetz}}]{PRB_count_edge_2023}%
  \BibitemOpen
  \bibfield  {author} {\bibinfo {author} {\bibfnamefont {U.}~\bibnamefont {Javed}}, \bibinfo {author} {\bibfnamefont {J.}~\bibnamefont {Marino}}, \bibinfo {author} {\bibfnamefont {V.}~\bibnamefont {Oganesyan}}, \ and\ \bibinfo {author} {\bibfnamefont {M.}~\bibnamefont {Kolodrubetz}},\ }\href {\doibase 10.1103/PhysRevB.108.L140301} {\bibfield  {journal} {\bibinfo  {journal} {Phys. Rev. B}\ }\textbf {\bibinfo {volume} {108}},\ \bibinfo {pages} {L140301} (\bibinfo {year} {2023})}\BibitemShut {NoStop}%
\bibitem [{\citenamefont {Sedlmayr}\ \emph {et~al.}(2023)\citenamefont {Sedlmayr}, \citenamefont {Cheraghi},\ and\ \citenamefont {Sedlmayr}}]{OTOC_inf_trap_PRB_2023}%
  \BibitemOpen
  \bibfield  {author} {\bibinfo {author} {\bibfnamefont {M.}~\bibnamefont {Sedlmayr}}, \bibinfo {author} {\bibfnamefont {H.}~\bibnamefont {Cheraghi}}, \ and\ \bibinfo {author} {\bibfnamefont {N.}~\bibnamefont {Sedlmayr}},\ }\href {\doibase 10.1103/PhysRevB.108.184303} {\bibfield  {journal} {\bibinfo  {journal} {Phys. Rev. B}\ }\textbf {\bibinfo {volume} {108}},\ \bibinfo {pages} {184303} (\bibinfo {year} {2023})}\BibitemShut {NoStop}%
\bibitem [{\citenamefont {Muruganandam}\ \emph {et~al.}(2024)\citenamefont {Muruganandam}, \citenamefont {Sajjan},\ and\ \citenamefont {Kais}}]{Muruganandam_2024}%
  \BibitemOpen
  \bibfield  {author} {\bibinfo {author} {\bibfnamefont {V.}~\bibnamefont {Muruganandam}}, \bibinfo {author} {\bibfnamefont {M.}~\bibnamefont {Sajjan}}, \ and\ \bibinfo {author} {\bibfnamefont {S.}~\bibnamefont {Kais}},\ }\href {\doibase 10.1088/1402-4896/ad7911} {\bibfield  {journal} {\bibinfo  {journal} {Physica Scripta}\ }\textbf {\bibinfo {volume} {99}},\ \bibinfo {pages} {105123} (\bibinfo {year} {2024})}\BibitemShut {NoStop}%
\bibitem [{\citenamefont {Sur}\ and\ \citenamefont {Sen}(2023)}]{Sur_2024}%
  \BibitemOpen
  \bibfield  {author} {\bibinfo {author} {\bibfnamefont {S.}~\bibnamefont {Sur}}\ and\ \bibinfo {author} {\bibfnamefont {D.}~\bibnamefont {Sen}},\ }\href {\doibase 10.1088/1361-648X/ad1363} {\bibfield  {journal} {\bibinfo  {journal} {Journal of Physics: Condensed Matter}\ }\textbf {\bibinfo {volume} {36}},\ \bibinfo {pages} {125402} (\bibinfo {year} {2023})}\BibitemShut {NoStop}%
\bibitem [{\citenamefont {Cavalcante}\ \emph {et~al.}(2025)\citenamefont {Cavalcante}, \citenamefont {Bonan\ifmmode~\mbox{\c{c}}\else \c{c}\fi{}a}, \citenamefont {Miranda},\ and\ \citenamefont {Deffner}}]{Moa_PRR_2025}%
  \BibitemOpen
  \bibfield  {author} {\bibinfo {author} {\bibfnamefont {M.~F.}\ \bibnamefont {Cavalcante}}, \bibinfo {author} {\bibfnamefont {M.~V.~S.}\ \bibnamefont {Bonan\ifmmode~\mbox{\c{c}}\else \c{c}\fi{}a}}, \bibinfo {author} {\bibfnamefont {E.}~\bibnamefont {Miranda}}, \ and\ \bibinfo {author} {\bibfnamefont {S.}~\bibnamefont {Deffner}},\ }\href {\doibase 10.1103/PhysRevResearch.7.L022027} {\bibfield  {journal} {\bibinfo  {journal} {Phys. Rev. Res.}\ }\textbf {\bibinfo {volume} {7}},\ \bibinfo {pages} {L022027} (\bibinfo {year} {2025})}\BibitemShut {NoStop}%
\bibitem [{\citenamefont {Sachdev}(1999)}]{sachdev1999quantum}%
  \BibitemOpen
  \bibfield  {author} {\bibinfo {author} {\bibfnamefont {S.}~\bibnamefont {Sachdev}},\ }\href {https://books.google.com/books?id=K_lhQgAACAAJ} {\emph {\bibinfo {title} {Quantum Phase Transitions}}}\ (\bibinfo  {publisher} {Cambridge University Press},\ \bibinfo {address} {Cambridge, UK},\ \bibinfo {year} {1999})\BibitemShut {NoStop}%
\bibitem [{\citenamefont {Talkner}\ \emph {et~al.}(2007)\citenamefont {Talkner}, \citenamefont {Lutz},\ and\ \citenamefont {H\"anggi}}]{PRE_work_2007}%
  \BibitemOpen
  \bibfield  {author} {\bibinfo {author} {\bibfnamefont {P.}~\bibnamefont {Talkner}}, \bibinfo {author} {\bibfnamefont {E.}~\bibnamefont {Lutz}}, \ and\ \bibinfo {author} {\bibfnamefont {P.}~\bibnamefont {H\"anggi}},\ }\href {\doibase 10.1103/PhysRevE.75.050102} {\bibfield  {journal} {\bibinfo  {journal} {Phys. Rev. E}\ }\textbf {\bibinfo {volume} {75}},\ \bibinfo {pages} {050102} (\bibinfo {year} {2007})}\BibitemShut {NoStop}%
\bibitem [{\citenamefont {Talkner}\ \emph {et~al.}(2008)\citenamefont {Talkner}, \citenamefont {H\"anggi},\ and\ \citenamefont {Morillo}}]{PRE_Talkner_2008}%
  \BibitemOpen
  \bibfield  {author} {\bibinfo {author} {\bibfnamefont {P.}~\bibnamefont {Talkner}}, \bibinfo {author} {\bibfnamefont {P.}~\bibnamefont {H\"anggi}}, \ and\ \bibinfo {author} {\bibfnamefont {M.}~\bibnamefont {Morillo}},\ }\href {\doibase 10.1103/PhysRevE.77.051131} {\bibfield  {journal} {\bibinfo  {journal} {Phys. Rev. E}\ }\textbf {\bibinfo {volume} {77}},\ \bibinfo {pages} {051131} (\bibinfo {year} {2008})}\BibitemShut {NoStop}%
\bibitem [{\citenamefont {Deffner}\ and\ \citenamefont {Campbell}(2019)}]{Deffner_Campbell_book}%
  \BibitemOpen
  \bibfield  {author} {\bibinfo {author} {\bibfnamefont {S.}~\bibnamefont {Deffner}}\ and\ \bibinfo {author} {\bibfnamefont {S.}~\bibnamefont {Campbell}},\ }\href {\doibase 10.1088/978-1-64327-657-1} {\emph {\bibinfo {title} {Quantum Thermodynamics}}}\ (\bibinfo  {publisher} {Morgan \& Claypool Publishers},\ \bibinfo {address} {San Rafael, CA},\ \bibinfo {year} {2019})\BibitemShut {NoStop}%
\bibitem [{\citenamefont {Gambassi}\ and\ \citenamefont {Silva}(2011)}]{gambassi2011}%
  \BibitemOpen
  \bibfield  {author} {\bibinfo {author} {\bibfnamefont {A.}~\bibnamefont {Gambassi}}\ and\ \bibinfo {author} {\bibfnamefont {A.}~\bibnamefont {Silva}},\ }\href@noop {} {} (\bibinfo {year} {2011}),\ \Eprint {http://arxiv.org/abs/1106.2671} {arXiv:1106.2671 [cond-mat.stat-mech]} \BibitemShut {NoStop}%
\bibitem [{\citenamefont {Smacchia}\ and\ \citenamefont {Silva}(2012)}]{PRL_Ale_2012}%
  \BibitemOpen
  \bibfield  {author} {\bibinfo {author} {\bibfnamefont {P.}~\bibnamefont {Smacchia}}\ and\ \bibinfo {author} {\bibfnamefont {A.}~\bibnamefont {Silva}},\ }\href {\doibase 10.1103/PhysRevLett.109.037202} {\bibfield  {journal} {\bibinfo  {journal} {Phys. Rev. Lett.}\ }\textbf {\bibinfo {volume} {109}},\ \bibinfo {pages} {037202} (\bibinfo {year} {2012})}\BibitemShut {NoStop}%
\bibitem [{\citenamefont {Sotiriadis}\ \emph {et~al.}(2013)\citenamefont {Sotiriadis}, \citenamefont {Gambassi},\ and\ \citenamefont {Silva}}]{PRE_Ale_2013}%
  \BibitemOpen
  \bibfield  {author} {\bibinfo {author} {\bibfnamefont {S.}~\bibnamefont {Sotiriadis}}, \bibinfo {author} {\bibfnamefont {A.}~\bibnamefont {Gambassi}}, \ and\ \bibinfo {author} {\bibfnamefont {A.}~\bibnamefont {Silva}},\ }\href {\doibase 10.1103/PhysRevE.87.052129} {\bibfield  {journal} {\bibinfo  {journal} {Phys. Rev. E}\ }\textbf {\bibinfo {volume} {87}},\ \bibinfo {pages} {052129} (\bibinfo {year} {2013})}\BibitemShut {NoStop}%
\bibitem [{\citenamefont {Smacchia}\ and\ \citenamefont {Silva}(2013)}]{PRE_Ale_2013_2}%
  \BibitemOpen
  \bibfield  {author} {\bibinfo {author} {\bibfnamefont {P.}~\bibnamefont {Smacchia}}\ and\ \bibinfo {author} {\bibfnamefont {A.}~\bibnamefont {Silva}},\ }\href {\doibase 10.1103/PhysRevE.88.042109} {\bibfield  {journal} {\bibinfo  {journal} {Phys. Rev. E}\ }\textbf {\bibinfo {volume} {88}},\ \bibinfo {pages} {042109} (\bibinfo {year} {2013})}\BibitemShut {NoStop}%
\bibitem [{\citenamefont {Russomanno}\ \emph {et~al.}(2016)\citenamefont {Russomanno}, \citenamefont {Sharma}, \citenamefont {Dutta},\ and\ \citenamefont {Santoro}}]{Russomanno_2016}%
  \BibitemOpen
  \bibfield  {author} {\bibinfo {author} {\bibfnamefont {A.}~\bibnamefont {Russomanno}}, \bibinfo {author} {\bibfnamefont {S.}~\bibnamefont {Sharma}}, \bibinfo {author} {\bibfnamefont {A.}~\bibnamefont {Dutta}}, \ and\ \bibinfo {author} {\bibfnamefont {G.~E.}\ \bibnamefont {Santoro}},\ }\href {\doibase 10.1088/1742-5468/2016/03/039901} {\bibfield  {journal} {\bibinfo  {journal} {Journal of Statistical Mechanics: Theory and Experiment}\ }\textbf {\bibinfo {volume} {2016}},\ \bibinfo {pages} {039901} (\bibinfo {year} {2016})}\BibitemShut {NoStop}%
\bibitem [{\citenamefont {Makki}\ \emph {et~al.}(2022)\citenamefont {Makki}, \citenamefont {Bandyopadhyay}, \citenamefont {Maity},\ and\ \citenamefont {Dutta}}]{PRB_Dutta_2022}%
  \BibitemOpen
  \bibfield  {author} {\bibinfo {author} {\bibfnamefont {A.~A.}\ \bibnamefont {Makki}}, \bibinfo {author} {\bibfnamefont {S.}~\bibnamefont {Bandyopadhyay}}, \bibinfo {author} {\bibfnamefont {S.}~\bibnamefont {Maity}}, \ and\ \bibinfo {author} {\bibfnamefont {A.}~\bibnamefont {Dutta}},\ }\href {\doibase 10.1103/PhysRevB.105.054301} {\bibfield  {journal} {\bibinfo  {journal} {Phys. Rev. B}\ }\textbf {\bibinfo {volume} {105}},\ \bibinfo {pages} {054301} (\bibinfo {year} {2022})}\BibitemShut {NoStop}%
\bibitem [{\citenamefont {Diehl}\ \emph {et~al.}(1983)\citenamefont {Diehl}, \citenamefont {Dietrich},\ and\ \citenamefont {Eisenriegler}}]{Diehl_PRB_1983}%
  \BibitemOpen
  \bibfield  {author} {\bibinfo {author} {\bibfnamefont {H.~W.}\ \bibnamefont {Diehl}}, \bibinfo {author} {\bibfnamefont {S.}~\bibnamefont {Dietrich}}, \ and\ \bibinfo {author} {\bibfnamefont {E.}~\bibnamefont {Eisenriegler}},\ }\href {\doibase 10.1103/PhysRevB.27.2937} {\bibfield  {journal} {\bibinfo  {journal} {Phys. Rev. B}\ }\textbf {\bibinfo {volume} {27}},\ \bibinfo {pages} {2937} (\bibinfo {year} {1983})}\BibitemShut {NoStop}%
\bibitem [{\citenamefont {Cardy}(1984)}]{CARDY1984514}%
  \BibitemOpen
  \bibfield  {author} {\bibinfo {author} {\bibfnamefont {J.~L.}\ \bibnamefont {Cardy}},\ }\href {\doibase https://doi.org/10.1016/0550-3213(84)90241-4} {\bibfield  {journal} {\bibinfo  {journal} {Nuclear Physics B}\ }\textbf {\bibinfo {volume} {240}},\ \bibinfo {pages} {514} (\bibinfo {year} {1984})}\BibitemShut {NoStop}%
\bibitem [{\citenamefont {Oshikawa}\ and\ \citenamefont {Affleck}(1997)}]{OSHIKAWA1997533}%
  \BibitemOpen
  \bibfield  {author} {\bibinfo {author} {\bibfnamefont {M.}~\bibnamefont {Oshikawa}}\ and\ \bibinfo {author} {\bibfnamefont {I.}~\bibnamefont {Affleck}},\ }\href {\doibase https://doi.org/10.1016/S0550-3213(97)00219-8} {\bibfield  {journal} {\bibinfo  {journal} {Nuclear Physics B}\ }\textbf {\bibinfo {volume} {495}},\ \bibinfo {pages} {533} (\bibinfo {year} {1997})}\BibitemShut {NoStop}%
\bibitem [{\citenamefont {Bernien}\ \emph {et~al.}(2017)\citenamefont {Bernien}, \citenamefont {Schwartz}, \citenamefont {Keesling}, \citenamefont {Levine}, \citenamefont {Omran}, \citenamefont {Pichler}, \citenamefont {Choi}, \citenamefont {Zibrov}, \citenamefont {Endres}, \citenamefont {Greiner}, \citenamefont {Vuleti{\'{c}}},\ and\ \citenamefont {Lukin}}]{Bernien2017}%
  \BibitemOpen
  \bibfield  {author} {\bibinfo {author} {\bibfnamefont {H.}~\bibnamefont {Bernien}}, \bibinfo {author} {\bibfnamefont {S.}~\bibnamefont {Schwartz}}, \bibinfo {author} {\bibfnamefont {A.}~\bibnamefont {Keesling}}, \bibinfo {author} {\bibfnamefont {H.}~\bibnamefont {Levine}}, \bibinfo {author} {\bibfnamefont {A.}~\bibnamefont {Omran}}, \bibinfo {author} {\bibfnamefont {H.}~\bibnamefont {Pichler}}, \bibinfo {author} {\bibfnamefont {S.}~\bibnamefont {Choi}}, \bibinfo {author} {\bibfnamefont {A.~S.}\ \bibnamefont {Zibrov}}, \bibinfo {author} {\bibfnamefont {M.}~\bibnamefont {Endres}}, \bibinfo {author} {\bibfnamefont {M.}~\bibnamefont {Greiner}}, \bibinfo {author} {\bibfnamefont {V.}~\bibnamefont {Vuleti{\'{c}}}}, \ and\ \bibinfo {author} {\bibfnamefont {M.~D.}\ \bibnamefont {Lukin}},\ }\href {\doibase 10.1038/nature24622} {\bibfield  {journal} {\bibinfo  {journal} {Nature}\ }\textbf {\bibinfo {volume} {551}},\ \bibinfo {pages} {579} (\bibinfo {year} {2017})}\BibitemShut {NoStop}%
\bibitem [{\citenamefont {Keesling}\ \emph {et~al.}(2019)\citenamefont {Keesling}, \citenamefont {Omran}, \citenamefont {Levine}, \citenamefont {Bernien}, \citenamefont {Pichler}, \citenamefont {Choi}, \citenamefont {Samajdar}, \citenamefont {Schwartz}, \citenamefont {Silvi}, \citenamefont {Sachdev}, \citenamefont {Zoller}, \citenamefont {Endres}, \citenamefont {Greiner}, \citenamefont {Vuleti{\'{c}}},\ and\ \citenamefont {Lukin}}]{Keesling2019}%
  \BibitemOpen
  \bibfield  {author} {\bibinfo {author} {\bibfnamefont {A.}~\bibnamefont {Keesling}}, \bibinfo {author} {\bibfnamefont {A.}~\bibnamefont {Omran}}, \bibinfo {author} {\bibfnamefont {H.}~\bibnamefont {Levine}}, \bibinfo {author} {\bibfnamefont {H.}~\bibnamefont {Bernien}}, \bibinfo {author} {\bibfnamefont {H.}~\bibnamefont {Pichler}}, \bibinfo {author} {\bibfnamefont {S.}~\bibnamefont {Choi}}, \bibinfo {author} {\bibfnamefont {R.}~\bibnamefont {Samajdar}}, \bibinfo {author} {\bibfnamefont {S.}~\bibnamefont {Schwartz}}, \bibinfo {author} {\bibfnamefont {P.}~\bibnamefont {Silvi}}, \bibinfo {author} {\bibfnamefont {S.}~\bibnamefont {Sachdev}}, \bibinfo {author} {\bibfnamefont {P.}~\bibnamefont {Zoller}}, \bibinfo {author} {\bibfnamefont {M.}~\bibnamefont {Endres}}, \bibinfo {author} {\bibfnamefont {M.}~\bibnamefont {Greiner}}, \bibinfo {author} {\bibfnamefont {V.}~\bibnamefont {Vuleti{\'{c}}}}, \ and\ \bibinfo {author} {\bibfnamefont {M.~D.}\ \bibnamefont {Lukin}},\ }\href {\doibase 10.1038/s41586-019-1070-1}
  {\bibfield  {journal} {\bibinfo  {journal} {Nature}\ }\textbf {\bibinfo {volume} {568}},\ \bibinfo {pages} {207} (\bibinfo {year} {2019})}\BibitemShut {NoStop}%
\bibitem [{\citenamefont {Altman}\ \emph {et~al.}(2021)\citenamefont {Altman}, \citenamefont {Brown}, \citenamefont {Carleo}, \citenamefont {Carr}, \citenamefont {Demler}, \citenamefont {Chin}, \citenamefont {DeMarco}, \citenamefont {Economou}, \citenamefont {Eriksson}, \citenamefont {Fu}, \citenamefont {Greiner}, \citenamefont {Hazzard}, \citenamefont {Hulet}, \citenamefont {Koll\'ar}, \citenamefont {Lev}, \citenamefont {Lukin}, \citenamefont {Ma}, \citenamefont {Mi}, \citenamefont {Misra}, \citenamefont {Monroe}, \citenamefont {Murch}, \citenamefont {Nazario}, \citenamefont {Ni}, \citenamefont {Potter}, \citenamefont {Roushan}, \citenamefont {Saffman}, \citenamefont {Schleier-Smith}, \citenamefont {Siddiqi}, \citenamefont {Simmonds}, \citenamefont {Singh}, \citenamefont {Spielman}, \citenamefont {Temme}, \citenamefont {Weiss}, \citenamefont {Vu\ifmmode \check{c}\else \v{c}\fi{}kovi\ifmmode~\acute{c}\else \'{c}\fi{}}, \citenamefont {Vuleti\ifmmode~\acute{c}\else \'{c}\fi{}}, \citenamefont {Ye},\ and\
  \citenamefont {Zwierlein}}]{PRXQuantum.2.017003}%
  \BibitemOpen
  \bibfield  {author} {\bibinfo {author} {\bibfnamefont {E.}~\bibnamefont {Altman}}, \bibinfo {author} {\bibfnamefont {K.~R.}\ \bibnamefont {Brown}}, \bibinfo {author} {\bibfnamefont {G.}~\bibnamefont {Carleo}}, \bibinfo {author} {\bibfnamefont {L.~D.}\ \bibnamefont {Carr}}, \bibinfo {author} {\bibfnamefont {E.}~\bibnamefont {Demler}}, \bibinfo {author} {\bibfnamefont {C.}~\bibnamefont {Chin}}, \bibinfo {author} {\bibfnamefont {B.}~\bibnamefont {DeMarco}}, \bibinfo {author} {\bibfnamefont {S.~E.}\ \bibnamefont {Economou}}, \bibinfo {author} {\bibfnamefont {M.~A.}\ \bibnamefont {Eriksson}}, \bibinfo {author} {\bibfnamefont {K.-M.~C.}\ \bibnamefont {Fu}}, \bibinfo {author} {\bibfnamefont {M.}~\bibnamefont {Greiner}}, \bibinfo {author} {\bibfnamefont {K.~R.}\ \bibnamefont {Hazzard}}, \bibinfo {author} {\bibfnamefont {R.~G.}\ \bibnamefont {Hulet}}, \bibinfo {author} {\bibfnamefont {A.~J.}\ \bibnamefont {Koll\'ar}}, \bibinfo {author} {\bibfnamefont {B.~L.}\ \bibnamefont {Lev}}, \bibinfo {author} {\bibfnamefont
  {M.~D.}\ \bibnamefont {Lukin}}, \bibinfo {author} {\bibfnamefont {R.}~\bibnamefont {Ma}}, \bibinfo {author} {\bibfnamefont {X.}~\bibnamefont {Mi}}, \bibinfo {author} {\bibfnamefont {S.}~\bibnamefont {Misra}}, \bibinfo {author} {\bibfnamefont {C.}~\bibnamefont {Monroe}}, \bibinfo {author} {\bibfnamefont {K.}~\bibnamefont {Murch}}, \bibinfo {author} {\bibfnamefont {Z.}~\bibnamefont {Nazario}}, \bibinfo {author} {\bibfnamefont {K.-K.}\ \bibnamefont {Ni}}, \bibinfo {author} {\bibfnamefont {A.~C.}\ \bibnamefont {Potter}}, \bibinfo {author} {\bibfnamefont {P.}~\bibnamefont {Roushan}}, \bibinfo {author} {\bibfnamefont {M.}~\bibnamefont {Saffman}}, \bibinfo {author} {\bibfnamefont {M.}~\bibnamefont {Schleier-Smith}}, \bibinfo {author} {\bibfnamefont {I.}~\bibnamefont {Siddiqi}}, \bibinfo {author} {\bibfnamefont {R.}~\bibnamefont {Simmonds}}, \bibinfo {author} {\bibfnamefont {M.}~\bibnamefont {Singh}}, \bibinfo {author} {\bibfnamefont {I.}~\bibnamefont {Spielman}}, \bibinfo {author} {\bibfnamefont {K.}~\bibnamefont
  {Temme}}, \bibinfo {author} {\bibfnamefont {D.~S.}\ \bibnamefont {Weiss}}, \bibinfo {author} {\bibfnamefont {J.}~\bibnamefont {Vu\ifmmode \check{c}\else \v{c}\fi{}kovi\ifmmode~\acute{c}\else \'{c}\fi{}}}, \bibinfo {author} {\bibfnamefont {V.}~\bibnamefont {Vuleti\ifmmode~\acute{c}\else \'{c}\fi{}}}, \bibinfo {author} {\bibfnamefont {J.}~\bibnamefont {Ye}}, \ and\ \bibinfo {author} {\bibfnamefont {M.}~\bibnamefont {Zwierlein}},\ }\href {\doibase 10.1103/PRXQuantum.2.017003} {\bibfield  {journal} {\bibinfo  {journal} {PRX Quantum}\ }\textbf {\bibinfo {volume} {2}},\ \bibinfo {pages} {017003} (\bibinfo {year} {2021})}\BibitemShut {NoStop}%
\bibitem [{\citenamefont {Zeiher}\ \emph {et~al.}(2017)\citenamefont {Zeiher}, \citenamefont {Choi}, \citenamefont {Rubio-Abadal}, \citenamefont {Pohl}, \citenamefont {van Bijnen}, \citenamefont {Bloch},\ and\ \citenamefont {Gross}}]{Rydb_PRX_2017}%
  \BibitemOpen
  \bibfield  {author} {\bibinfo {author} {\bibfnamefont {J.}~\bibnamefont {Zeiher}}, \bibinfo {author} {\bibfnamefont {J.-y.}\ \bibnamefont {Choi}}, \bibinfo {author} {\bibfnamefont {A.}~\bibnamefont {Rubio-Abadal}}, \bibinfo {author} {\bibfnamefont {T.}~\bibnamefont {Pohl}}, \bibinfo {author} {\bibfnamefont {R.}~\bibnamefont {van Bijnen}}, \bibinfo {author} {\bibfnamefont {I.}~\bibnamefont {Bloch}}, \ and\ \bibinfo {author} {\bibfnamefont {C.}~\bibnamefont {Gross}},\ }\href {\doibase 10.1103/PhysRevX.7.041063} {\bibfield  {journal} {\bibinfo  {journal} {Phys. Rev. X}\ }\textbf {\bibinfo {volume} {7}},\ \bibinfo {pages} {041063} (\bibinfo {year} {2017})}\BibitemShut {NoStop}%
\bibitem [{\citenamefont {Rylands}\ \emph {et~al.}(2020)\citenamefont {Rylands}, \citenamefont {Rozenbaum}, \citenamefont {Galitski},\ and\ \citenamefont {Konik}}]{PhysRevLett.124.155302}%
  \BibitemOpen
  \bibfield  {author} {\bibinfo {author} {\bibfnamefont {C.}~\bibnamefont {Rylands}}, \bibinfo {author} {\bibfnamefont {E.~B.}\ \bibnamefont {Rozenbaum}}, \bibinfo {author} {\bibfnamefont {V.}~\bibnamefont {Galitski}}, \ and\ \bibinfo {author} {\bibfnamefont {R.}~\bibnamefont {Konik}},\ }\href {\doibase 10.1103/PhysRevLett.124.155302} {\bibfield  {journal} {\bibinfo  {journal} {Phys. Rev. Lett.}\ }\textbf {\bibinfo {volume} {124}},\ \bibinfo {pages} {155302} (\bibinfo {year} {2020})}\BibitemShut {NoStop}%
\end{thebibliography}%

\end{document}